\documentclass[aps,twocolumn,reprint]{revtex4-1}

\draft 

\usepackage{wrapfig}
\usepackage[]{graphicx,xcolor}
\usepackage{tabularx}
\usepackage{booktabs}
\usepackage{textcomp}
\usepackage{amsmath}
\usepackage{amssymb}
\usepackage[]{graphics}
\usepackage{amssymb}
\usepackage{amsmath}
\usepackage{color}
\usepackage{ifpdf}
\usepackage{lipsum}
\usepackage{siunitx}
\usepackage{braket}
\usepackage{soul}

\graphicspath{{figs/}} 
\newcommand{\executeiffilenewer}[3]{%
\ifnum\pdfstrcmp{\pdffilemoddate{#1}}%
{\pdffilemoddate{#2}} > 0 {\immediate\write18{#3}}\fi}

\ifpdf\usepackage{epstopdf}\fi

\begin{document}


\title{Wide band enhancement of the transverse magneto-optical Kerr effect in magnetite-based plasmonic crystals}

\author{S.~A.~Dyakov}
\email[]{e-mail: s.dyakov@skoltech.ru}
\affiliation{Skolkovo Institute of Science and Technology, 143025 Moscow Region, Russia}

\author{L.~Klompmaker}
\affiliation{Experimentelle Physik 2, Technische Universit\"{a}t Dortmund, 44221 Dortmund, Germany}

\author{F.~Spitzer}
\affiliation{Experimentelle Physik 2, Technische Universit\"{a}t Dortmund, 44221 Dortmund, Germany}

\author{I.~M.~Fradkin}
\affiliation{Skolkovo Institute of Science and Technology, 143025 Moscow Region, Russia}

\author{E. Yalcin}
\affiliation{Experimentelle Physik 2, Technische Universit\"{a}t Dortmund, 44221 Dortmund, Germany}

\author{I.~A. Akimov}
\affiliation{Experimentelle Physik 2, Technische Universit\"{a}t Dortmund, 44221 Dortmund, Germany}
\affiliation{Ioffe Institute, 194021 St. Petersburg,  Russia}

\author{D.~A.~Yavsin}
\affiliation{Ioffe Institute, 194021 St. Petersburg,  Russia}

\author{S.~I.~Pavlov}
\affiliation{Ioffe Institute, 194021 St. Petersburg,  Russia}

\author{S.~Y.~Verbin}
\affiliation{Spin Optics Laboratory, Saint Petersburg State University, 198504 St. Petersburg, Russia}

\author{S.~G.~Tikhodeev}
\affiliation{A.~M.~Prokhorov General Physics Institute, RAS, Vavilova 38, Moscow, Russia}
\affiliation{Faculty of Physics, Lomonosov Moscow State University, 119991 Moscow, Russia}

\author{N.~A.~Gippius}
\affiliation{Skolkovo Institute of Science and Technology, 143025 Moscow Region, Russia}

\author{A.~B.~Pevtsov}
\affiliation{Ioffe Institute, 194021 St. Petersburg,  Russia}

\author{M.~Bayer}
\affiliation{Experimentelle Physik 2, Technische Universit\"{a}t Dortmund, 44221 Dortmund, Germany}
\affiliation{Ioffe Institute, 194021 St. Petersburg,  Russia}


\date{\today}
\begin{abstract}
The transverse magneto-optical Kerr effect (TMOKE) in magnetite-based magneto-plasmonic crystals is studied experimentally and theoretically. We analyse angle-resolved TMOKE spectra from two types of structures where noble metallic stripes are incorporated inside a thin magnetite film or located on top of a homogeneous film. A multiple wide band enhancement of the TMOKE signal in transmission is demonstrated. The complex dielectric permittivity and gyration are experimentally determined using the ellipsometry technique as well as Faraday rotation and ellipticity measurements. The obtained parameters are used in rigorous coupled wave analysis (RCWA) calculations for studying the optical resonances. Our RCWA calculations of transmittance and TMOKE are in good agreement with the experimental data. The role of guiding and plasmonic modes in the TMOKE enhancement is revealed. We demonstrate that the TMOKE provides rich information about the studied optical resonances.
\end{abstract}
\maketitle
\section{Introduction}
Magneto-optical effects in magnetic nanostructure materials are currently attracting a great deal of attention because they provide a possibility to control the intensity of the reflected and transmitted light. This is the foundation of the potential of the magneto-optical effects for data storage \cite{betzig1992near}, for use in optical isolation systems \cite{gauthier1986simple, Iwamura1978, ishida2017amorphous, shoji2008magneto}, in various magnetic \cite{lenz1990review, zu2012magneto, didosyan2003magneto} and biological sensors \cite{diaz2017enhanced, sepulveda2006highly, park2009magneto,regatos2011suitable}, and in optical filtering \cite{rotondaro2015generalized, keaveney2018optimized}. Magneto-optical effects can also be used for the realization of ultrafast optical switches in nanophotonic circuits where instead of a slowly varying external magnetic field short radio-frequency or optical pulses are applied to govern the magnetization dynamics in magnetic media \cite{bossini2016}. 

One of the most common and prominent intensity effects is the transverse magneto-optical Kerr effect (TMOKE) which is defined by the relative change $\delta$ of the reflected or transmitted intensities $I$ for the two opposite directions of an in-plane magnetization $M$ at the interface between two materials:
\begin{equation}
\delta = 2\frac{I(M)-I(-M)}{I(M)+I(-M)},
\label{eqdelta}
\end{equation}
with the incident plane of light perpendicular to the magnetization direction. TMOKE has been actively studied in ferromagnetic metals \cite{krinchik1968magneto, martin1965optical, zvezdin1997, grunin2010surface} and has been used for the investigation of the energy structure in ferromagnetic systems. In planar structures, e.g.\ homogeneous films consisting of conventional magnetic materials, the TMOKE has values less than $\delta=$\,10$^{-3}$ which substantially limits its applicability. Yet, TMOKE has an important feature, namely it is determined by the magnetic properties of the interface and sample geometry and therefore can be used for the control of light at the nanoscale. Recent advances in nanotechnology allow one to synthesize magnetic nanostructures where the magnitude of TMOKE is significantly increased in the vicinity of optical resonances \cite{maksymov2014transverse, barsukova2017magneto, amanollahi2018wide, PhysRevB.99.085440, giron2017giant, satoru2017magneto}. In particular, TMOKE enhancement has been demonstrated in magnetoplasmonic crystals \cite{belotelov2011enhanced, akimov2012hybrid, gonzalez2008plasmonic, pohl2013tuning, clavero2010magnetic, newman2008magneto, dyakov2018transverse, borovkova2018tmoke} and magnetoplasmonic nanoantennas \cite{maksymov2016magneto, chen2011plasmonic,valente2015magneto, loughran2018enhancing} where plasmonic resonances come into play. 

\begin{figure*}[t!]
\centering
\includegraphics[width=0.8\textwidth]{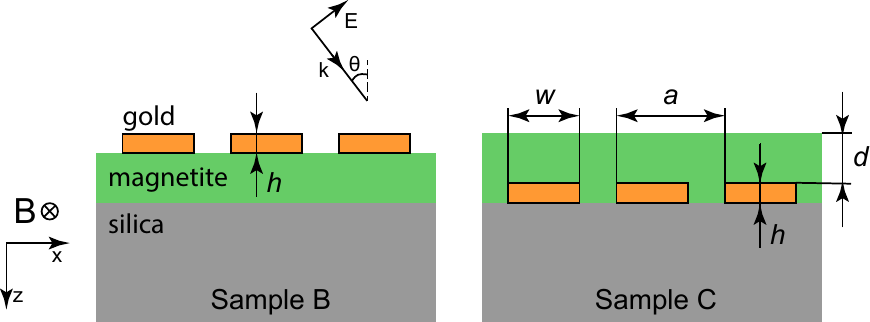}
\caption{(Color online) Sketch of the samples used in the experiments and simulations. In all samples the magnetite film thickness $d=100$\,nm, the grating period $a=580$\,nm, the gold stripes width $w=400$\,nm.}
\label{sample}
\end{figure*} 

 \begin{figure}[t!]
\centering
\includegraphics[width=1\columnwidth]{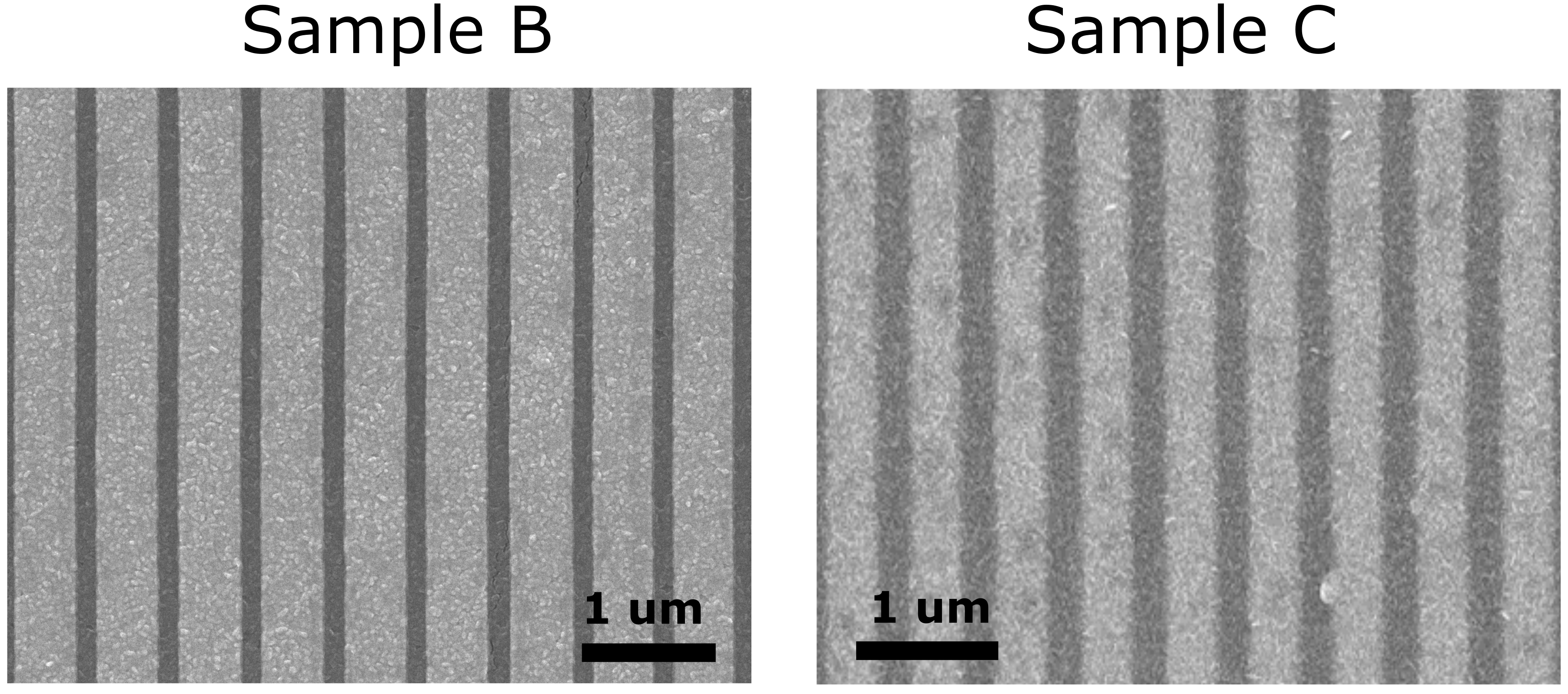}
\caption{(Color online) Scanning electron microscopy images of samples B and C.}
\label{photosample}
\end{figure} 

Of particular interest is the TMOKE in transmission obtained in structures which lack of mirror symmetry. Moreover, in some applications a finite transparency is required which excludes purely metallic structures based on ferromagnetic metals. Progress has been achieved in hybrid plasmonic structures where the combination of a noble metal and a magnetic dielectric with low optical losses results in a significant enhancement of TMOKE in the vicinity of surface plasmon polariton resonances \cite{belotelov2011enhanced, akimov2012hybrid}. TMOKE magnitudes up to $\delta=$\,15\% have been reported in bismuth substituted iron garnets (BIG) covered with gold gratings \cite{pohl2013tuning, kreilkamp2013waveguide}. 

In Ref.\,\onlinecite{kreilkamp2013waveguide} a magnetoplasmonic system with a thin film of bismuth iron garnet (BIG) and a grating of narrow gold stripes was investigated. It was demonstrated that in such a system the TMOKE enhancement is due to the excitation of waveguide-plasmon polaritons \cite{Christ2003b, PhysRevB.93.205413}, quasiparticles formed due to the strong interaction between a guided resonance mode in the BIG waveguide and localized surface plasmons in the narrow gold stripes. The investigation of different systems of magnetic materials with noble metallic structures promises new interesting functionalities, especially concerning TMOKE in transmission.
 


In this work we study magnetoplasmonic crystals consisting of a magnetite film with gold nanostripe arrays. The great interest in TMOKE properties of magnetite is due to the fact that it is the most magnetic of all the naturally-occurring minerals on Earth. Magnetite belongs to the class of ferrimagnetics; its magnetic properties were realized already since ancient times. Magnetite offers a large variety of applications, from biomedical to environmental.

In this article, we investigate what types of plasmonic modes appear in such a system and which values of TMOKE enhancement one can expect in them. To the best of our knowledge, there are no publications describing the TMOKE properties of magnetoplasmonic systems containing magnetite. Therefore, we believe that this study can be interesting both from practical and fundamental viewpoints.

\section {Description of samples}
In our theoretical and experimental studies we consider the two samples shown in Fig. 1, each consisting of a magnetite film with a gold stripe grating (Fig.\,\ref{sample}). In both samples the magnetite film thickness is $d=100$\,nm, the grating period is $a=580$\,nm, the gold stripe width is $w=400$\,nm and the gold thickness is $h=40$\,nm. We also consider planar samples without periodicity which we use as a reference. The geometry of these planar samples will be explained hereinafter.

The magnetic films containing Fe$_3$O$_4$/a-Fe nanoparticle complexes were synthesized with the laser electrodispersion technique onto a quartz substrate with subsequent annealing in vacuum (10$^{-4}$\,Pa) at $T = 300^\circ$C for 1\,h. Single-crystal Fe$_3$O$_4$ was used as initial target \cite{melekh2016nanostructured}. X-ray diffraction and electron microscopy studies showed that the average size of the magnetic nanoparticles in the films was 6--10\,nm and these particles have the crystal structure of magnetite \cite{melekh2016nanostructured}. The coercive force and the saturation magnetization of the synthesized nanostructured films were as large as $\sim$660\,Oe and $\sim$520\,emu/cm$^3$, respectively. These values are considerably higher than the corresponding parameters of polycrystalline Fe$_3$O$_4$ films.


200$\times$200\,$\mu$m arrays of gold nanostripes were created by lift-off electron beam lithography as follows: 300\,nm thick positive e-beam resist PMMA 950K (Allresist, GmbH) was spin-coated on the substrate and e-beam lithography was performed. After developing the e-beam pattern, 40\,nm gold with a 5\,nm titanium adhesive layer were deposited by thermal evaporation and the lift-off process was performed leading to the gold stripes. E-beam lithography and SEM characterization was carried out by a electron microscope JSM 7001F (JEOL, Japan), equipped with the e-beam lithography system 'Nanomaker' (Interface Ltd, Russia). SEM images of samples B and C are shown in Fig.\,\ref{photosample}. Because of electron scattering in the magnetite film, the visible proportions of the gold stripes of sample C are distorted in the SEM image. Consequently, the gold stripes of the same width in samples B and C look different in their SEM images.


\begin{figure}[t!]
\centering
\includegraphics[width=0.8\columnwidth]{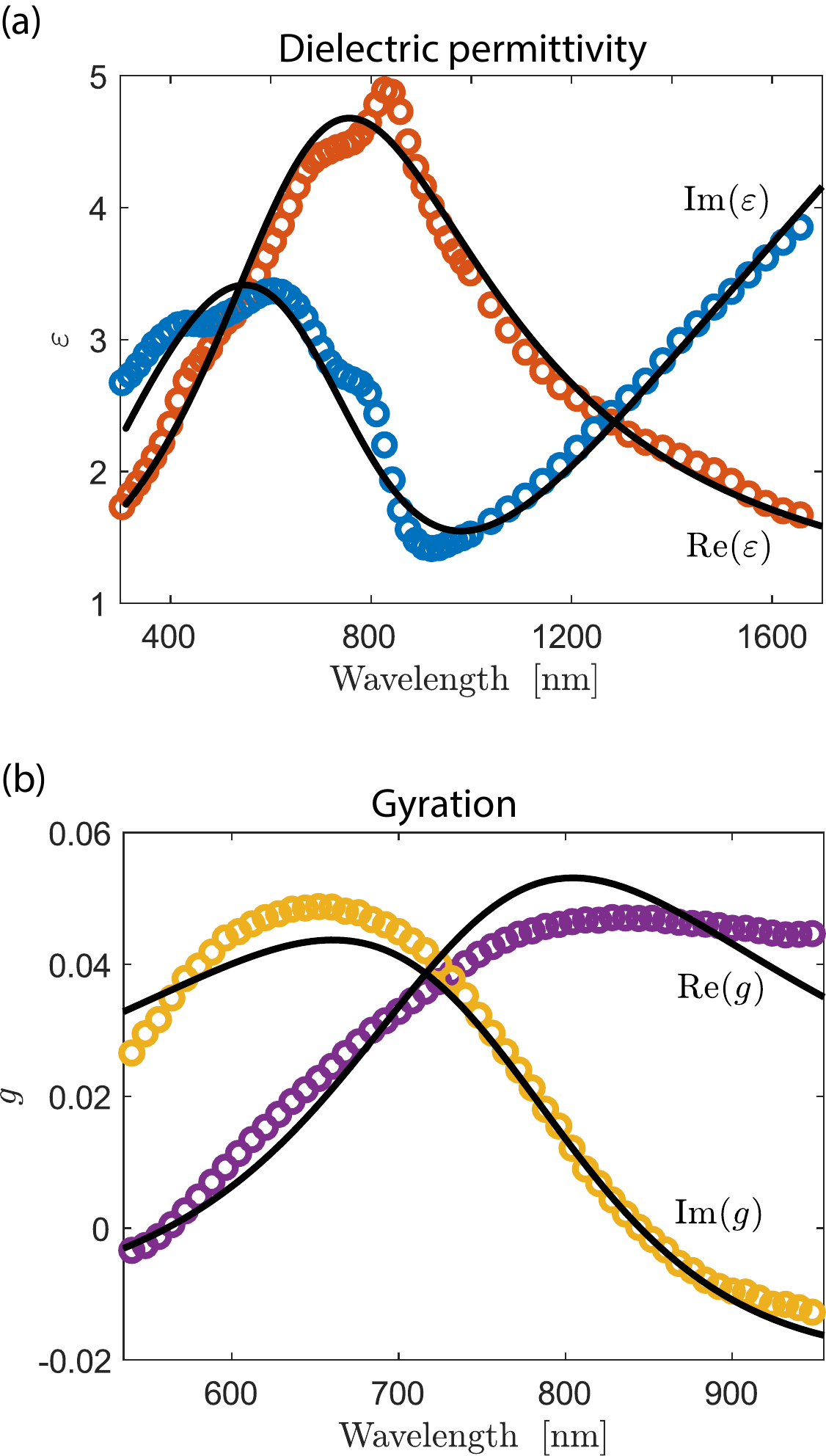}
\caption{(a) Real and imaginary parts of the dielectric permittivity of magnetite evaluated from ellipsometry measurements (dots) and our analytic model (black curves).  (b) Real and imaginary parts of the $z$-component of the gyration vector evaluated from the spectral dependences of Faraday rotation and ellipticity in Faraday geometry, respectively, measured on a homogeneous film. Black curves represent the analytic model for gyration.}
\label{epsgy}
\end{figure} 

\section{Optical constants of magnetite}
The optical properties of the magnetized magnetite can be described macroscopically by the non-diagonal dielectric permittivity tensor which in the linear approximation for the magnetization has the form    
\begin{equation}
\hat{\varepsilon}=
\begin{bmatrix}
    \varepsilon      & -ig_z & ig_y \\
    ig_z       & \varepsilon & -ig_x \\
    -ig_y      & ig_x & \varepsilon 
\end{bmatrix},
\end{equation}
where $\varepsilon$ is the dielectric permittivity of magnetite in the absence of magnetization. The diagonal components of the dielectric tensor are all equal because there is no anisotropy in nanocrystalline films of Fe$_{3}$O$_{4}$. The complex gyration vector $\vec{g} = (g_x, g_y, g_z)$ is proportional to the magnetization $\vec{g} = \alpha \vec{M}$ where the proportionality factor $\alpha$ does not depend on the magnetization direction. The magnetic permeability $\mu$ is assumed to be unit. 


The complex dielectric permittivity $\varepsilon$ was determined by the standard ellipsometry technique and is shown in Fig.\,\ref{epsgy}(a). One can note from  Fig.\,\ref{epsgy}(a) that magnetite has rather high internal optical losses in the visible range. 

For the theoretical consideration of magneto-optical effects, we also determined the complex gyration $g_z$ (see Fig.\,\ref{epsgy}(b)) by measuring the Faraday rotation and ellipticity where the direction of light propagation was parallel to the magnetic field direction and perpendicular to the magnetic film plane. These measurements were performed in a magnetic field of about 0.6\,T which is above the saturation level of the synthesized magnetite films. Since the parameter $\alpha$ does not depend on the magnetization direction, we assumed that $g_x=g_y=g_z$.

In spite of the fact that our magnetic films are non-uniform and consist of Fe$_3$O$_4$/a-Fe nanoparticle complexes, the obtained dielectric permittivity and gyration are close to the literature data measured for monocrystalline and epitaxial films \cite{fontijn1997optical, bobo2001magnetic}.

To calculate the eigenmodes we need to know dielectric functions of materials at complex frequencies. For such dispersive material as magnetite, this is not a simple problem to restore its dielectric permittivity beyond the real axis. The most convenient way, in this case, is the construction of an analytical fitting function. For this purpose, we used the program developed in Ref.\, \onlinecite{sehmi2017optimizing}, which fits the permittivity by the Drude model and a number of Lorentz poles.

We considered magnetite in a simple model with a single Lorentz pole: 
\begin{equation}
    \varepsilon(\omega) = \varepsilon_\infty+\frac{i\sigma_\mathrm{D}}{\omega}-\frac{i\sigma_\mathrm{D}}{\omega+i\gamma_\mathrm{D}}+\frac{i\sigma_\mathrm{L}}{\omega -\Omega_\mathrm{L}}+\frac{i\sigma^*_\mathrm{L}}{\omega +\Omega^*_\mathrm{L}}
\end{equation}    

Although the gyration of magnetite is not used for calculation of the eigenmodes, we fitted it with a single Lorentz pole for a proper extrapolation of the experimental data into the frequency band of interest:
\begin{equation}
    g(\omega) = \varepsilon_\infty+\frac{i\sigma_\mathrm{L}}{\omega -\Omega_\mathrm{L}}+\frac{i\sigma^*_\mathrm{L}}{\omega +\Omega^*_\mathrm{L}}.
\end{equation}   

The dielectric permittivity of gold given by the Johnson and Christy data \cite{johnson1972optical} was fitted just by a Drude term. Optimized parameters of the models are given in table \ref{tab:parameters} and the comparison between experimental data and analytical results is depicted in Fig. \ref{epsgy}.

\begin{table}[h]
  \begin{tabular}{ c|c|c|c }
	     			& Au	           & $\mathrm{Fe}_3 \mathrm{O}_4$ & $g$	\\
    \hline
 \rule{0pt}{3ex}
     $\varepsilon_\infty$	& 11.0310	& 1.3211 & -0.0128	\\
      $\gamma ({\rm eV})$	& 0.0928	& 19.0790 & -	\\
      $\sigma ({\rm eV})$	& 932.3400	& 4.2403 & -	\\
   \hline
  \rule{0pt}{3ex}
    $\Omega_L' ({\rm eV})$	& -	& 1.4198 &  1.5706 \\
     $ \Omega_L'' ({\rm eV})$	& -	& -0.7463 & -0.3650	\\
     $ \sigma_L' ({\rm eV})$	& -	& 2.6652 & 0.0232 	\\
      $\sigma_L'' ({\rm eV})$	& -	& -1.6837 & 0.0036	\\
  \end{tabular}
  \caption{Optimized model parameters for gold, magnetite and its gyration $g$.}
  \label{tab:parameters}
\end{table}

\section{Theoretical method}
To calculate the reflection and transmission spectra we used a Fourier modal method in the scattering matrix form \cite{Tikhodeev2002b}, also known as rigorous coupled wave analysis (RCWA) \cite{moharam1995formulation}. In order to achieve a better convergence with respect to the number of plane waves, we employ the Li's factorization rules \cite{li1996use}. In our RCWA simulations, we used 51 plane waves. The eigenmodes of structures are calculated by finding the poles of the scattering matrix \cite{Gippius2005c}. 

\section{Optical resonances in magnetoplasmonic crystals with magnetite}

\begin{figure*}[t!]
\centering
\includegraphics[width=0.8\textwidth]{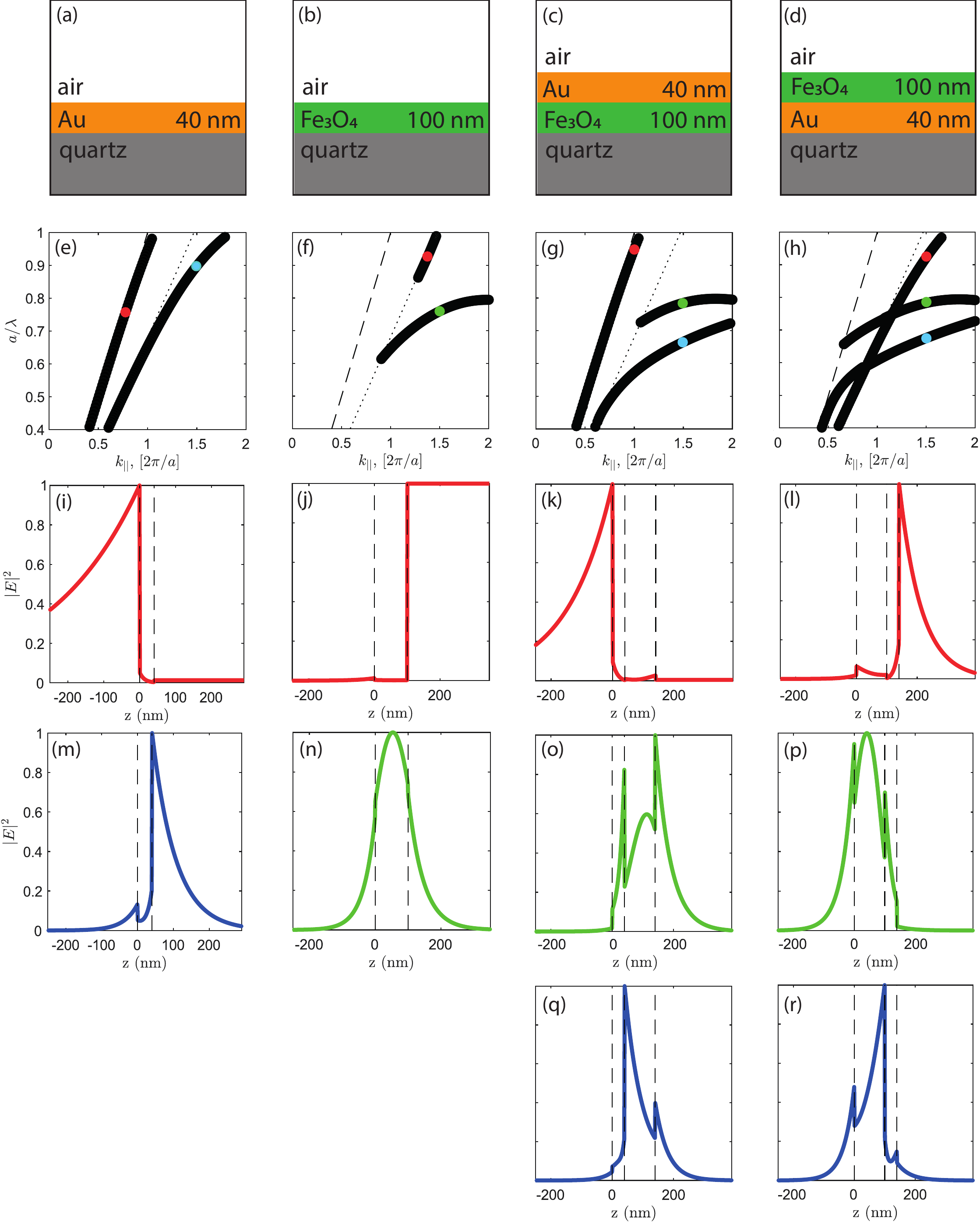}
\caption{(Color online) Sketches of planar structures (a--d), their eigenmodes (e--h) and normalized electric field intensity profiles (i--r). In panels (e--h) dashed and dotted lines show the air and silica light line, respectively. The mode profiles in (i--r) are calculated at points specified by red, blue and green circles on the dispersion curves in (e--h). Colors are used to differentiate the field profiles without further physical meaning. Dashed vertical lines in (i--r) show the interfaces between layers. The $z=0$ position corresponds always to the interface with air.}
\label{ELA}
\end{figure*}

To understand the features of light interaction with our magnetoplasmonic systems we start with the theoretical analysis of eigenmodes. We begin with the simplest structures and then gradually increase their complexity as it is shown in Fig. 4 (a-c). The behaviour of eigenmodes of simple homogeneous structures is important because, as we will demonstrate, it helps explaining the more complex modes picture of the structures with periodic gratings.

The dispersion and intensity profiles of the eigenmodes of a 40\,nm thick homogeneous gold film on a silica substrate are shown in Fig.\,\ref{ELA}(e,i,m). These modes are surface plasmon polaritons (SPPs) along the air/gold and gold/silica interfaces and are mixed with each other due to their close spatial localization. 

Next, the eigenmodes of a 100\,nm thick magnetite film on a silica substrate are described in Fig.\,\ref{ELA}(f,j,n). Since the refractive index of magnetite is larger than the refractive index of silica, these modes represent the guided resonances of the magnetite film waveguide. The upper branch is TM polarized, while the lower branch is TE polarized. Due to the boundary conditions of the electric displacement vector and the high refractive index ratio of magnetite and silica, the guided mode in TM polarization almost perfectly avoids the magnetite film and is localized mainly in the substrate (Fig.\,\ref{ELA}(j)). In contrast, the guided mode in TE polarization is confined within the waveguide (Fig.\,\ref{ELA}(n)).

Further, we merge the gold and magnetite films, first by placing the gold film above the magnetite film (Fig.\,\ref{ELA}(c)). As a result we obtain three hybrid modes shown in Fig.\,\ref{ELA}(g). The dispersion curve of the upper mode reminds to the gold/air SPP mode of the plain gold film on silica, i.e. it is close to the air light line and hence its field profile is mainly localized in air. The dispersions of the other modes are below the silica light line; they are localized both in the magnetite film and the substrate. 

Finally, we consider the case where the gold film is below the magnetite film (Fig.\,\ref{ELA}(d)). In such a structure one also can observe three hybrid modes (Fig.\,\ref{ELA}(h)). One of the modes is close to the silica light line and represents the gold/silica SPP. The two other modes are below the air light line, they decay in air but can propagate in the substrate. 

We would like to emphasize that the modes of structures composed of gold and magnetite are hybrid and the plasmonic and photonic contributions to the different modes differ. 

\begin{figure*}[t!]
\centering
\includegraphics[width=1\textwidth]{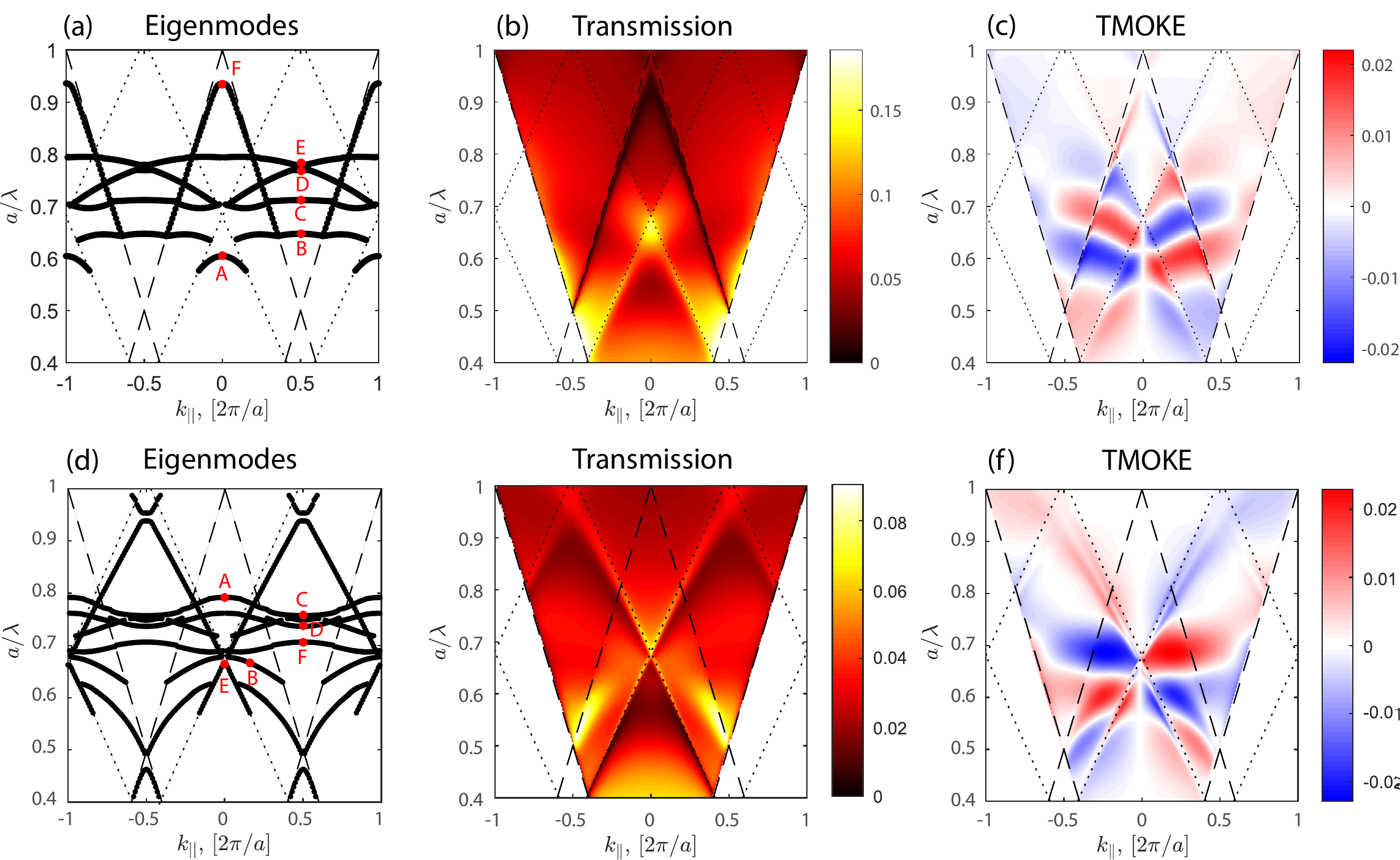}
\caption{(Color online) Panels (a) and (d) show the eigenmodes of the samples B and C, respectively. The dispersions of the real part of the eigenenergies are shown by lines. Calculated transmission coefficients (b), (e) and TMOKE (c), (f) of the samples B and C as a function of in-plane wavevector $k_x$ and $a/\lambda$. Red dots in panels (a) and (d) denote points where the electric field distribution is calculated.}
\label{TMOKE}
\end{figure*}

\begin{figure*}[t!]
\centering
\includegraphics[width=0.85\textwidth]{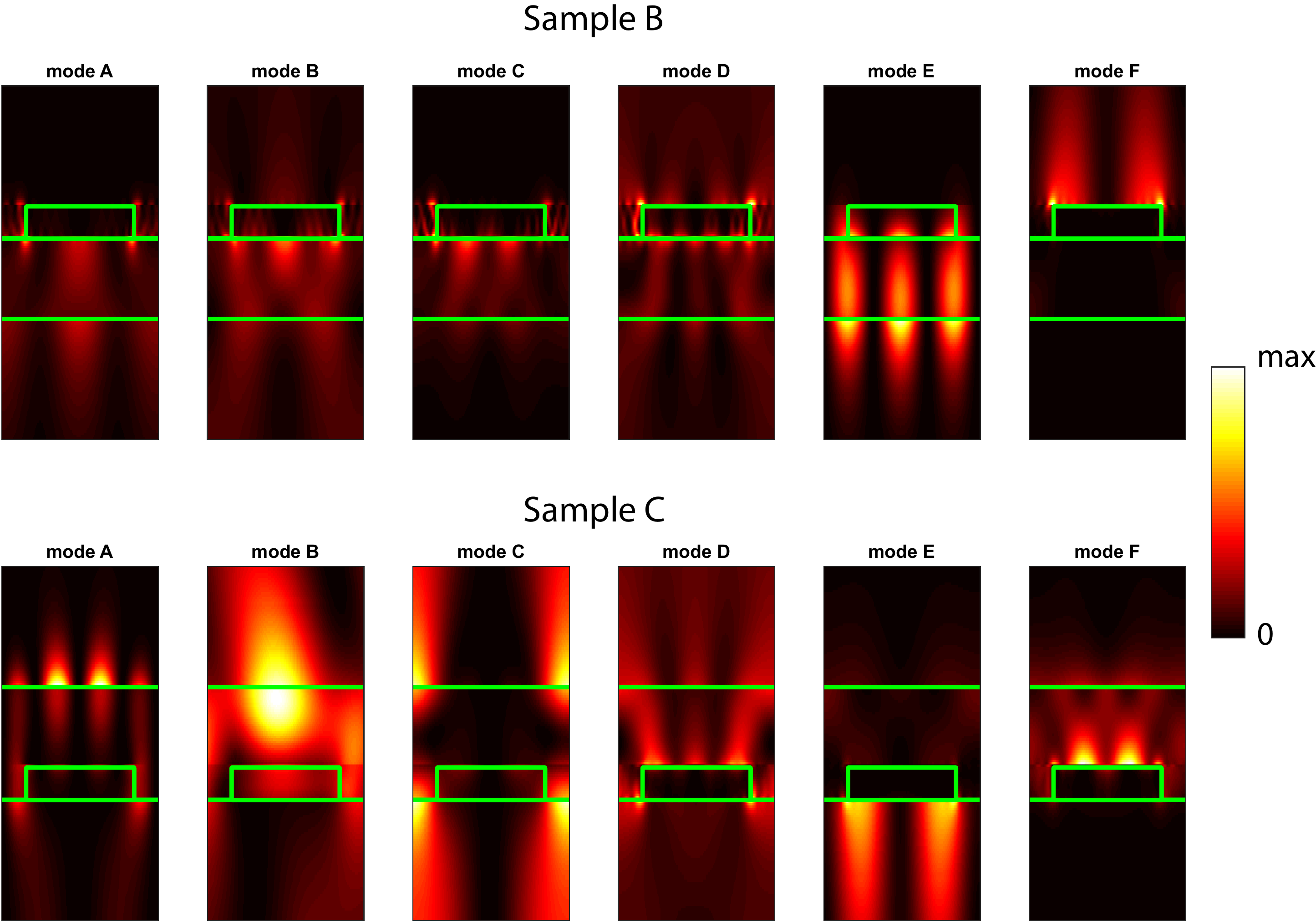}
\caption{(Color online) Electric field intensity distribution in samples B and C at the frequencies and wavevectors specified by the red dots in Fig.\,\ref{TMOKE}(a),(d). Colorscale is shown on the right.}
\label{fields}
\end{figure*}

All modes described so far are below the air light line (dashed line in Figs.\,\ref{ELA}(e--h)) and, hence, they do not appear in far field transmission or reflection spectra due to the momentum conservation law. The presence of a periodic gold grating in the samples B and C folds these modes into the first Brillouin zone leading to Fano resonances and making them visible in the far field \cite{Tikhodeev2002b, dyakov2017quasiguided}. In periodic systems like ours, at the high-symmetry points of the Brillouin zone ($k_{\parallel}=0, \frac{\pi}{a}$, etc) the modes degeneracy is partially lifted which gives rise to families of modes (see, for example \cite{Sakoda2001a,Tikhodeev2002b,dyakov2018magnetic}). In addition to that, localized surface plasmon modes can appear on individual gold stripes \cite{Christ2004}. All of these modes interact with each other creating the eigenmode dispersions shown in Fig.\,\ref{TMOKE}. The field distributions of some of these modes are shown in Fig.\,\ref{fields}.

\begin{figure}[t!]
\centering
\includegraphics[width=1\columnwidth]{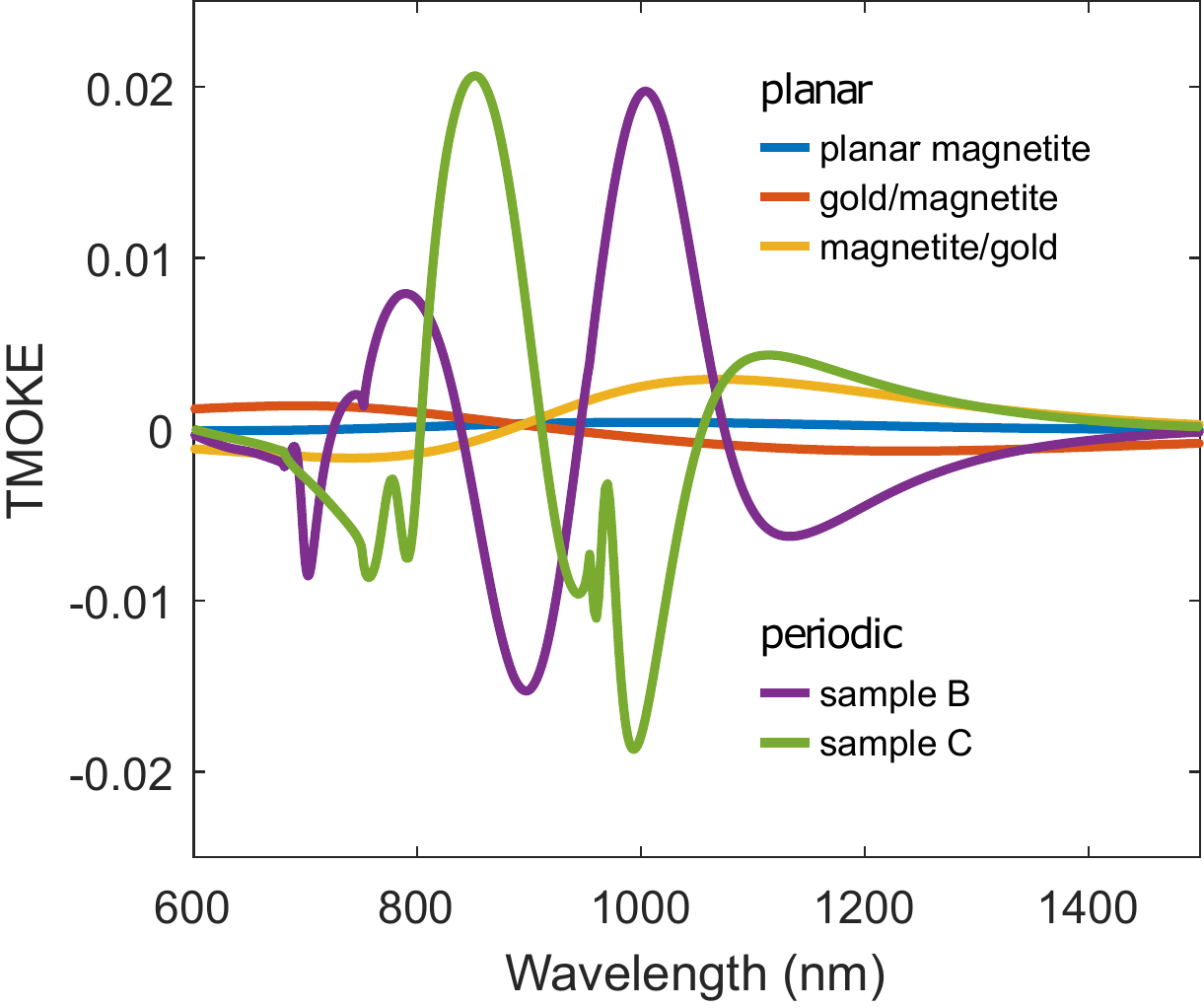}
\caption{(Color online) TMOKE spectra calculated for the three planar samples from Fig.\,\ref{ELA}(b--d), the two magnetoplasmonic systems B and C, and the one periodic magnetite slab waveguide without gold. Thicknesses of periodic and nonperiodic parts are 40\,nm and 100\,nm, respectively, period is 580\,nm, air trenches width is 180\,nm.}
\label{spectracomp}
\end{figure} 


The resulting photon quasimomentum $k_{\parallel}$ and energy $a/\lambda$ dependences of transmission and TMOKE for the samples B and C are shown \footnote{Please note that the model of sample C used in simulations includes air trenches with 40 nm depth and 180 nm width.} in Fig.\,\ref{TMOKE}(b,c,e,f). The spectral positions of the transmission maxima and minima are close to some of the dispersion curves shown in the Fig.\,\ref{TMOKE}(c,d). Also, one can see that due to the high absorption in magnetite, the transmission and TMOKE spectra do not have sharp resonance peaks which are usually observed for low absorbing periodic slab waveguides or magnetoplasmonic crystals \cite{tikhodeev2005waveguide, kreilkamp2013waveguide, belotelov2011enhanced}. As a result, the spectral features in the TMOKE spectra are wide. 


It is worth noting that the TMOKE spectra of the samples B and C appear to be rather informative as they exhibit color changes (the white stripes) which often, but not always, can be associated with a particular resonance that is not clearly visible in the transmission spectra. To demonstrate this, we consider the example of a line with a normalized Lorentzian shape belonging to an arbitrary resonance:
\begin{equation}
    L_0(\omega)=\frac{I_0\gamma^2}{(\omega-\omega_0)^2+\gamma^2},
\end{equation}
where $\omega_0$ is the position of the maximum, $\gamma$ and $I_0$ are the parameters of the Lorentzian. Suppose that under the action of an external magnetic field, the spectral line is shifted. We consider two Lorentzian functions 
\begin{equation}
    L_{1,2}(\omega)=\frac{I_0\gamma^2}{(\omega-\omega_0\pm\Delta \omega)^2+\gamma^2},
\end{equation}
where the spectral shift $0 < \Delta \omega \ll \gamma$. It can easily be shown that 
\begin{equation}
    L_{1,2}(\omega)\approx L_0(\omega)\left[1\mp\frac{2\Delta \omega(\omega-\omega_0)L_0(\omega)}{I_0\gamma^2}\right].
\end{equation}
The resultant TMOKE signal reads as
\begin{equation}
    \delta = 2\frac{L_2(\omega)-L_1(\omega)}{L_2(\omega)+L_1(\omega)} = \frac{4\Delta \omega(\omega-\omega_0)L_0(\omega)}{I_0\gamma^2}.
    \label{eq:deltaTheo}
\end{equation}
This expression immediately gives the change of the sign of the TMOKE response at the resonance frequency $\omega_0$. That is why TMOKE spectra in general provide better contrast for resonance observation than transmission spectra. Indeed, in the TMOKE spectrum shown in Fig.\,\ref{TMOKE}(c), the lowest white stripe at $(a/\lambda)\simeq 0.55$ indicates the existence of an eigenmode below the intersection of the folded silica light lines. As can be seen in Fig.\,\ref{TMOKE}(a) this mode indeed exists (red dot), however it hardly can be seen in the corresponding transmission spectra (Fig.\,\ref{TMOKE}(b)).

The above discussion reveals two mechanisms of modes hybridization in the magnetoplasmonic crystals B and C. The first mechanism is the coupling between the surface plasmon polaritons and the guided modes due to the their close spatial localization. The second mechanism is the interaction of these already hybrid resonances with the photon continuum, which leads to the appearance of Fano-type resonances in the optical transmission and reflection spectra of such structures. 

In order to understand which mechanism is responsible for the TMOKE enhancement we compare the TMOKE spectra calculated for the three planar samples from Fig.\,\ref{ELA}(b--d), and the two magnetoplasmonic systems B and C. One can see from Fig.\,\ref{spectracomp} that only the magnetoplasmonic systems B and C have the TMOKE signal enhanced with respect to the planar magnetite film on silica. This proves that both hybridization mechanisms are needed for observing strong TMOKE signal. 

\section{Experimental demonstration of the TMOKE enhancement}

\begin{figure}[t!]
\centering
\includegraphics[width=0.9\columnwidth]{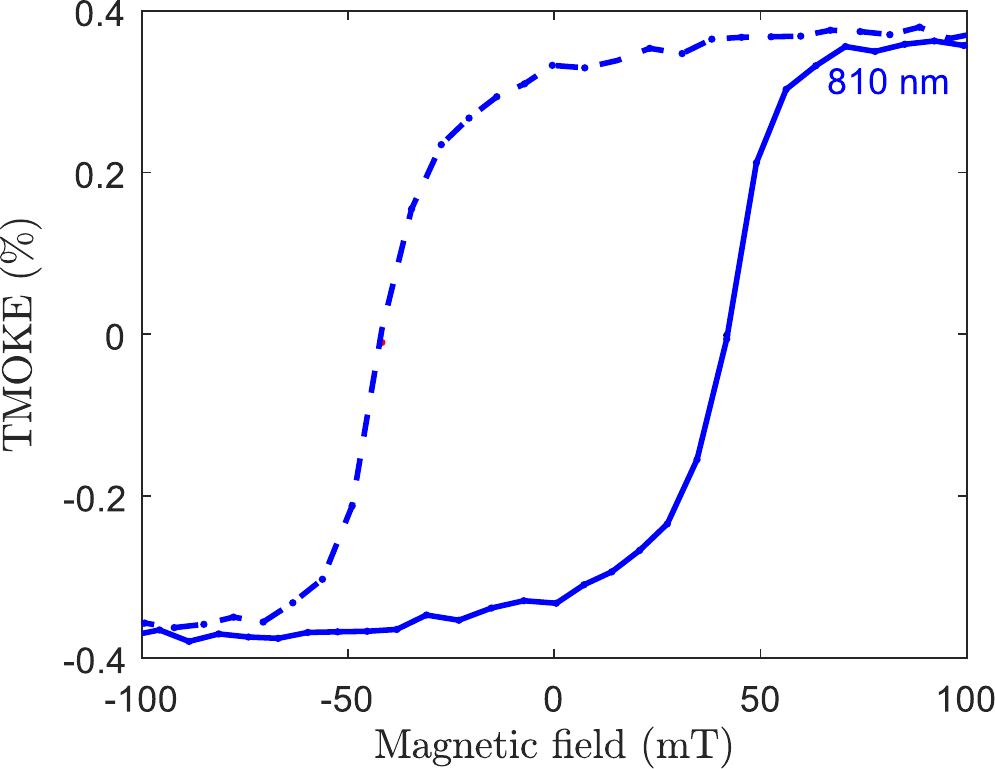}
\caption{(Color online) Magnetic field dependences of the TMOKE for a structure similar to sample B measured at 23$^\circ$ angle of light incidence at $\lambda = \SI{810}{nm}$. Solid line corresponds to Eq. \ref{eqdeltaHyst}, dashed line to switched indices.}
\label{hysteresis}
\end{figure} 
\begin{figure*}[t!!]
\centering
\includegraphics[width=0.9\textwidth]{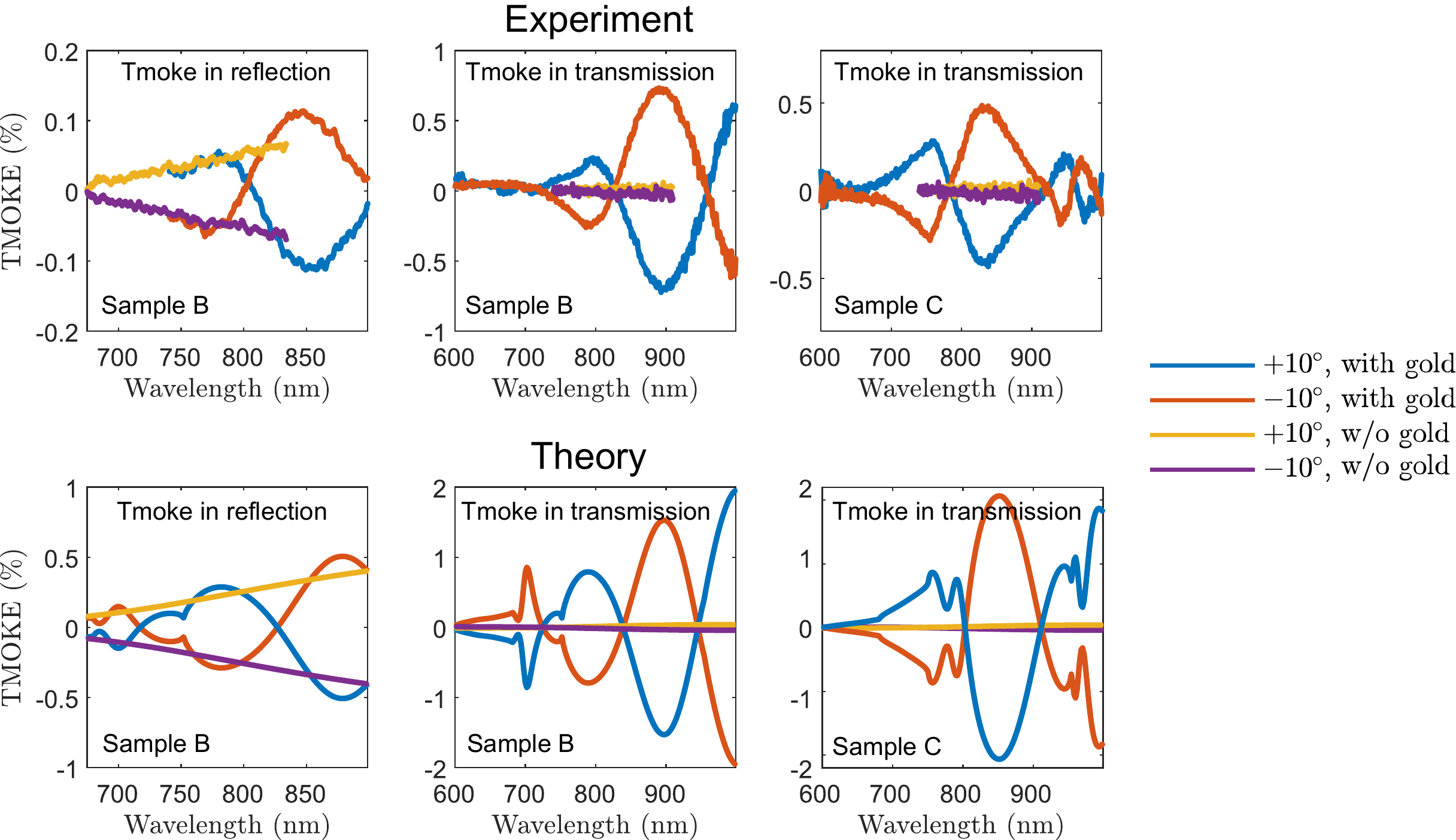}
\caption{(Color online) Experimental (a--c) and theoretical (d--f) spectra of TMOKE in reflection (a,d) and transmission (b,c,e,f) for the samples  B and C. Data are shown for incidence angles of +10$^\circ$ and -10$^\circ$.}
\label{spectraU}
\end{figure*} 
\begin{figure*}[t!]
\centering
\includegraphics[width=1\textwidth]{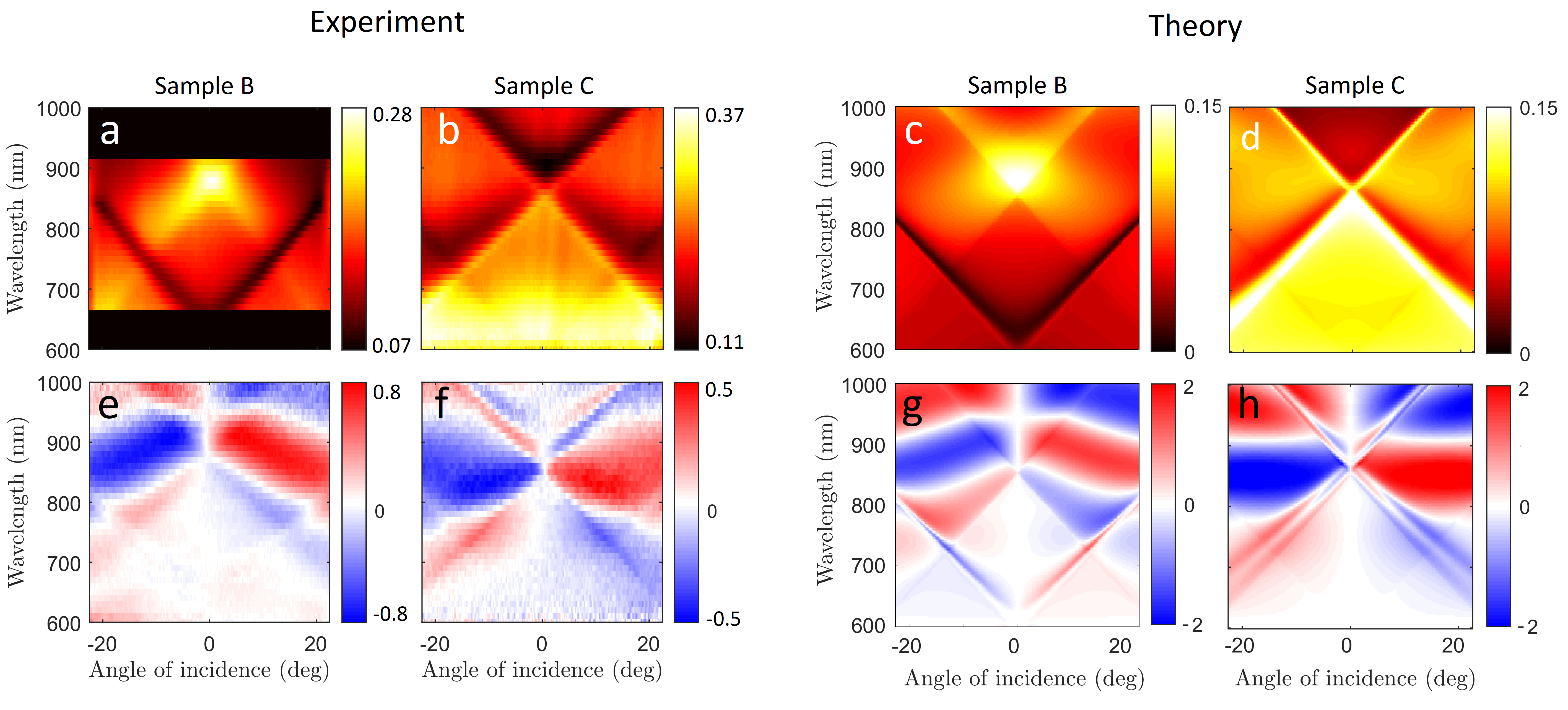}
\caption{(Color online) Theoretical and experimental wavelength and angle dependences of transmission (a--d) and TMOKE in transmission (e--h) for the samples B and C in p-polarization.}
\label{TMOKEcomparizon}
\end{figure*}

In our experimental studies the samples are positioned between the ferrite cores of an electromagnet, with the external magnetic field varying from 0 up to \SI{0.6}{T} oriented in-plane of the magnetic film and perpendicular to the incidence plane.
Angle and wavelength resolved reflectivity and transmission spectra were measured at a temperature of 300\,K using a tungsten halogen lamp, which illuminates the sample with $p$-polarized light.

The dependence of the TMOKE magnitude on the external magnetic field strength can be used as an indicator for the magnetization curve of the sample. 
These measurements were conducted for one particular incidence angle of \SI[separate-uncertainty = true]{23 \pm 1}{\degree} by rotating the sample around the axis parallel to the direction of the transverse magnetic field. 
An achromatic doublet lens was used to focus the p-polarized light onto the grating area and a second one to focus the transmitted light onto the spectrometer slit and the attached CCD.
Because magnetite is known to show magnetic hysteresis, the magnetic field was swept between \SI{-105}{mT} and \SI{105}{mT} in forward and backward direction in steps of \SI{7}{mT}, reaching saturation magnetization of the sample at the peak values. 
At each strength of the magnetic field the light intensity was detected.
In accordance with Eq. \ref{eqdelta} the parameter $\delta$ was calculated by comparing the transmitted intensities $I_{x}(\pm B)$ at opposite magnetic field directions $\pm B$, where the subscript $x$ indicates the direction of the magnetic field scan ($f$ being forward, $b$ backward)
\begin{equation}
    \delta = 2\frac{I_\mathrm{f}(+B) - I_\mathrm{b}(-B)}{I_\mathrm{f}(+B) + I_\mathrm{b}(-B)}\\
    \label{eqdeltaHyst}
\end{equation}
and with switched subscripts to get a complete hysteresis curve.

Fig. \,\ref{hysteresis} shows this magnetic field dependence of the TMOKE for the wavelength $\lambda = \SI{810}{nm}$ of the transmitted light for a magnetite structure with a gold grating on top, similar to sample B.
The solid line is calculated according to Eq. \ref{eqdeltaHyst} and the dashed one represents the dependence with switched indices in the equation.
Saturation of the TMOKE and thus of the sample magnetisation $M$ already takes place at a magnetic field of \SI{100}{mT}, while hysteresis behavior is identified for lower magnetic fields. 
The shape of the hysteresis curve of the TMOKE is the same for different wavelengths, always saturating at \SI{100}{mT}, but its magnitude and sign change in the vicinity of a resonance as shown by Eq. \ref{eq:deltaTheo}.  
We notice that the saturation level for the transverse Kerr effect geometry (Voigt) is about one order of magnitude smaller than that for the polar Kerr effect in Faraday geometry \cite{bobo2001magnetic}.
This is due to the appearance of a demagnetizing field which depends on the orientation of the applied magnetic field with respect to the sample surface \cite{chikazumi2009physics}.
For the polar Kerr effect the demagnetizing field is maximal because the magnetic field is oriented along the magnetite films' hard axis and along the easy axis for the TMOKE geometry, leading to the need of weaker magnetic field application in order to reach the saturation magnetization. All further measurements were conducted at saturating external magnetic fields.

To measure TMOKE at different angles we use a Fourier imaging spectroscopy setup \cite{richard2005spontaneous}.
A microscope objective with an N.A. of 0.4 is used both to focus the incident light onto the sample and to collect the reflected light in backscattering geometry, resulting in an angular range of -23$^\circ$ to +23$^\circ$ that can be covered in experiment.
For transmission measurements an identical microscope objective is used behind the sample to collimate the transmitted light.
In both cases a telescope consisting of two achromatic doublets maps the collimated light onto the imaging spectrometer slit and the CCD behind it, providing a spectral resolution of 0.6\,nm and an angular resolution of about 0.4$^\circ$.
To deduce the TMOKE parameter $\delta$ according to Eq. \ref{eqdelta} the reflected or transmitted intensity is measured for opposite magnetic field directions at saturated sample magnetization.
By switching the magnetic field direction multiple times and repeating the measurement each time, fluctuations of the lamps intensity were eliminated and the signal to noise ratio was increased.

Fig.\,\ref{spectraU} shows in the top row spectrally resolved TMOKE measurements in reflection (sample B) and transmission geometry (samples B and C), each comparing the effect from the plain magnetite film (yellow and purple) to that from the plasmonic structures (blue and orange).
All measurements shown select an angle of \SI{\pm 10}{\degree} from the Fourier imaging spectra.
In the bottom row we show the results of the theoretical simulations for comparison.
The TMOKE in reflection geometry (left) reaches a maximum value of \SI{0.1}{\percent} both for plain magnetite and combined with the plasmonic grating.
While the TMOKE magnitude from the plain magnetite film shows a monotonic rise with increasing wavelength, the plasmonic signal changes sign once at \SI{800}{nm} and likely also a second time around \SI{900}{nm}, which is slightly out of the measured range.
The maximum TMOKE occurs around \SI{850}{nm}. Compared to reflection we observe a general enhancement of the plasmonic TMOKE magnitude in transmission (center), reaching \SI{0.75}{\percent} at \SI{890}{nm}.
The general shape of the signal is the same with all features moved to slightly higher wavelengths, e.g. the two zero crossings now occur at \SI{825}{nm} and \SI{958}{nm}.
On sample C (right), with the gold grating located at the interface between magnetite and substrate, the resonances shift and the spectral dependence of the TMOKE signal changes compared to sample B, revealing more features.
We observe the maximum TMOKE signal of \SI{0.45}{\percent} around \SI{835}{nm}, now with three zero crossings occurring around \SI{786}{nm}, \SI{920}{nm} and \SI{958}{nm}.
On both samples the TMOKE in transmission from the plain magnetite film is weaker compared to reflection, only reaching about \SI{0.03}{\percent} while maintaining the general monotonically growing shape.
The enhancement of the TMOKE signal due to the grating structure is observed only in transmission. 
We attribute this effect to the excitation of the hybrid modes discussed above.
In all samples the peak to peak width of the TMOKE response is in the order of \SI{100}{nm} and thus spectrally broad.
This can be attributed mainly to the broad plasmonic and waveguide resonances seen in the transmission spectra.

The two-dimensional angle resolved TMOKE spectra in transmission are shown in Fig. \ref{TMOKEcomparizon} on the left in comparison to the theoretical simulations shown on the right.
These Fourier plots present a more intuitive way to follow the dispersion of resonant modes. 
The top row shows the transmission through the respective sample, going from black through red and yellow to white for increasing transmission.
Here the transmission is defined as the transmitted intensity through the sample with gold grating normalized to the intensity without the grating on the sample, thus showing the influence of the grating structure on the transmitted light and removing influences of optics or the lamp spectrum.
Both for sample B and C resonant features due to the grating structure are observed as minima or maxima in the transmission spectra.
On sample B the feature with decreased transmission in the p-polarized measurements can be attributed to a plasmon resonance, as this feature is not visible in measurements with s-polarized light. 
Due to the grating period of \SI{580}{nm} the resonance can be attributed to the air-gold plasmon.
The plasmon at the gold-magnetite interface is expected at lower wavelength outside of the measured range due to the higher refractive index of magnetite. A feature with increased transmission crossing the plasmonic resonance is also visible in s-polarized measurements with slightly shifted resonance frequency due to different boundary conditions.
It is not visible without the grating structure though, identifying it as waveguide mode inside the magnetite film.

Looking at the same points in the TMOKE measurement below, the plasmonic resonance is vaguely visible, but not as sign change of a strong TMOKE signal that is usually connected to a plasmonic resonance in a magnetoplasmonic system influenced by a magnetic field.
This confirms that this plasmonic resonance is located at the interface between the gold grating and the surrounding air and thus is influenced by the magnetic field only weakly through the spatially close magnetic magnetite film and a hybridization of the two crossing modes.
The region close to the bright resonance shows the strongest TMOKE signal, reaching $\delta = \SI{\pm 0.75}{\percent}$ in the wide colored area between \SI{800}{nm} and \SI{950}{nm} and showing the usual sign change but with lower magnitude on the other side towards lower wavelengths.
Close to the crossing point with the plasmon resonance an influence on the waveguide mode is visible, which deviates from the expected straight path towards higher wavelengths due to stronger hybridization of the two modes.
This further confirms the interaction between the magnetic waveguide and the non-magnetic plasmonic resonance.
Such an interaction was shown in previous studies on iron garnet films supporting plasmonic resonances \cite{pohl2013tuning}.
With increasing distance from the waveguide mode the TMOKE magnitude decreases.
Measurements in s-polarization show TMOKE neither from the resonances nor the non-resonant regions, as expected. 
The theoretical simulations reproduce these features well.
The same resonances are visible, leading to the same TMOKE signal.
The hybridization between the two modes can be seen notably well in the simulated TMOKE spectrum, showing a comparably weak signal for the plasmonic resonance which is highest at the crossing point with the magnetic waveguide mode.

Sample C shows a similar transmission and TMOKE spectrum (Fig\,\ref{TMOKEcomparizon}(c), again showing the waveguide mode. 
The waveguide mode shows an X-shape in the transmission spectrum, which separates regions of higher and lower transmission.
This leads to a region of low transmission at high wavelengths and small angles and one with high transmission for small angles and wavelengths.
Because the gold grating is located inside the magnetite film the air-gold plasmon is visible only as a faint remnant due to the rather thin magnetite cover. Also the plasmon resonance between gold and the silica substrate is not inside the measured wavelength range. 

In the TMOKE spectrum the highest TMOKE is reached in the region between \SI{800}{nm} and \SI{900}{nm} with \SI{0.45}{\percent}.
The waveguide mode can be identified here, too, replicating the X-shape seen in the transmission spectrum: 
The upper section of the cross ($\lambda > \SI{850}{nm}$) is seen as a white line with no TMOKE between areas of opposite TMOKE (red and blue) and the same trend as the waveguide mode in the transmission spectrum. 
For wavelengths below the crossing point of the waveguide mode ($\lambda < \SI{850}{nm}$) in the transmission spectrum, the TMOKE's white line is not perfectly following the straight waveguide resonance anymore, mainly close to the crossing point with the remnant of the air-gold plasmon around \SI{730}{nm}. 
As already seen on sample B, due to the spectral overlap of the TMOKE signals of the two resonances the TMOKE signal's white line is also shifted towards higher wavelengths, but less strongly for sample C.
A local maximum of TMOKE is visible at the waveguide mode position seen in the transmission spectrum.
The mentioned plasmonic resonance between air and gold is only weakly visible in the TMOKE spectrum, both in experiment and theory, indicating a weaker influence by the magnetic materials. 
Moving the gold grating into the magnetite film thus proves as a good strategy to still allow the excitation of waveguide modes but diminish the influence of the plasmon resonances. Again the simulations show a picture similar to the experimental data, namely inhibiting the plasmonic resonance and its influence on the strong waveguide mode in the transmission and TMOKE spectra.

\section{Conclusions}
In conclusion, we have studied the TMOKE in synthesized structures with magnetite and gold nanostripes.
We have demonstrated the multiple wide-band enhancement of the TMOKE response in magnetoplasmonic crystals compared to the bare magnetite film. 
The effect takes place in transmission only, where enhancement by more than an order of magnitude is observed.
We have shown that this effect is due to the hybridization of the guided modes of the magnetite film and the plasmonic modes of the gold grating. 
The hybridization between the modes can be optimized by proper choice of the geometry of the structure and its parameters, as demonstrated for the case of the air-metal plasmon resonance. 
Since the easy axis of the magnetic films is located in the plane of the structure, the TMOKE magnitude saturates already in small magnetic fields of about 100\,mT and shows clear hysteresis behaviour, allowing one to perform switching of the light intensity between two states. 
This can be useful for future investigations of these structures and possible applications, because the wide-band TMOKE response is robust against changes of wavelength and incidence angle of the light in these regions.

\vspace{1cm}
\section{Acknowledgements}
The authors thank the Deutsche Forschungsgemeinschaft (DFG) within the framework of the International Collaborative Research Centre (ICRC) TRR 160 (Project C5). E-beam lithography and SEM characterization were carried out in the Joint Research Center 'Materials science and characterization in advanced technology' with financial support by the Ministry of Education and Science of the Russian Federation (Agreement 14.621.21.0013, 28.18.2017, id RFMEFI62117X0018). The theoretical part of this work was supported by the Russian Science Foundation (Grant No. 16-12-10538).


\begin{thebibliography}{57}%
\makeatletter
\providecommand \@ifxundefined [1]{%
 \@ifx{#1\undefined}
}%
\providecommand \@ifnum [1]{%
 \ifnum #1\expandafter \@firstoftwo
 \else \expandafter \@secondoftwo
 \fi
}%
\providecommand \@ifx [1]{%
 \ifx #1\expandafter \@firstoftwo
 \else \expandafter \@secondoftwo
 \fi
}%
\providecommand \natexlab [1]{#1}%
\providecommand \enquote  [1]{``#1''}%
\providecommand \bibnamefont  [1]{#1}%
\providecommand \bibfnamefont [1]{#1}%
\providecommand \citenamefont [1]{#1}%
\providecommand \href@noop [0]{\@secondoftwo}%
\providecommand \href [0]{\begingroup \@sanitize@url \@href}%
\providecommand \@href[1]{\@@startlink{#1}\@@href}%
\providecommand \@@href[1]{\endgroup#1\@@endlink}%
\providecommand \@sanitize@url [0]{\catcode `\\12\catcode `\$12\catcode
  `\&12\catcode `\#12\catcode `\^12\catcode `\_12\catcode `\%12\relax}%
\providecommand \@@startlink[1]{}%
\providecommand \@@endlink[0]{}%
\providecommand \url  [0]{\begingroup\@sanitize@url \@url }%
\providecommand \@url [1]{\endgroup\@href {#1}{\urlprefix }}%
\providecommand \urlprefix  [0]{URL }%
\providecommand \Eprint [0]{\href }%
\providecommand \doibase [0]{http://dx.doi.org/}%
\providecommand \selectlanguage [0]{\@gobble}%
\providecommand \bibinfo  [0]{\@secondoftwo}%
\providecommand \bibfield  [0]{\@secondoftwo}%
\providecommand \translation [1]{[#1]}%
\providecommand \BibitemOpen [0]{}%
\providecommand \bibitemStop [0]{}%
\providecommand \bibitemNoStop [0]{.\EOS\space}%
\providecommand \EOS [0]{\spacefactor3000\relax}%
\providecommand \BibitemShut  [1]{\csname bibitem#1\endcsname}%
\let\auto@bib@innerbib\@empty
\bibitem [{\citenamefont {Betzig}\ \emph {et~al.}(1992)\citenamefont {Betzig},
  \citenamefont {Trautman}, \citenamefont {Wolfe}, \citenamefont {Gyorgy},
  \citenamefont {Finn}, \citenamefont {Kryder},\ and\ \citenamefont
  {Chang}}]{betzig1992near}%
  \BibitemOpen
  \bibfield  {author} {\bibinfo {author} {\bibfnamefont {E.}~\bibnamefont
  {Betzig}}, \bibinfo {author} {\bibfnamefont {J.}~\bibnamefont {Trautman}},
  \bibinfo {author} {\bibfnamefont {R.}~\bibnamefont {Wolfe}}, \bibinfo
  {author} {\bibfnamefont {E.}~\bibnamefont {Gyorgy}}, \bibinfo {author}
  {\bibfnamefont {P.}~\bibnamefont {Finn}}, \bibinfo {author} {\bibfnamefont
  {M.}~\bibnamefont {Kryder}}, \ and\ \bibinfo {author} {\bibfnamefont {C.-H.}\
  \bibnamefont {Chang}},\ }\href@noop {} {\bibfield  {journal} {\bibinfo
  {journal} {Applied Physics Letters}\ }\textbf {\bibinfo {volume} {61}},\
  \bibinfo {pages} {142} (\bibinfo {year} {1992})}\BibitemShut {NoStop}%
\bibitem [{\citenamefont {Gauthier}\ \emph {et~al.}(1986)\citenamefont
  {Gauthier}, \citenamefont {Narum},\ and\ \citenamefont
  {Boyd}}]{gauthier1986simple}%
  \BibitemOpen
  \bibfield  {author} {\bibinfo {author} {\bibfnamefont {D.~J.}\ \bibnamefont
  {Gauthier}}, \bibinfo {author} {\bibfnamefont {P.}~\bibnamefont {Narum}}, \
  and\ \bibinfo {author} {\bibfnamefont {R.~W.}\ \bibnamefont {Boyd}},\
  }\href@noop {} {\bibfield  {journal} {\bibinfo  {journal} {Optics letters}\
  }\textbf {\bibinfo {volume} {11}},\ \bibinfo {pages} {623} (\bibinfo {year}
  {1986})}\BibitemShut {NoStop}%
\bibitem [{\citenamefont {Iwamura}\ \emph {et~al.}(1978)\citenamefont
  {Iwamura}, \citenamefont {Hayashi},\ and\ \citenamefont
  {Iwasaki}}]{Iwamura1978}%
  \BibitemOpen
  \bibfield  {author} {\bibinfo {author} {\bibfnamefont {H.}~\bibnamefont
  {Iwamura}}, \bibinfo {author} {\bibfnamefont {S.}~\bibnamefont {Hayashi}}, \
  and\ \bibinfo {author} {\bibfnamefont {H.}~\bibnamefont {Iwasaki}},\ }\href
  {\doibase 10.1007/BF00620305} {\bibfield  {journal} {\bibinfo  {journal}
  {Optical and Quantum Electronics}\ }\textbf {\bibinfo {volume} {10}},\
  \bibinfo {pages} {393} (\bibinfo {year} {1978})}\BibitemShut {NoStop}%
\bibitem [{\citenamefont {Ishida}\ \emph {et~al.}(2017)\citenamefont {Ishida},
  \citenamefont {Miura}, \citenamefont {Shoji}, \citenamefont {Yokoi},
  \citenamefont {Mizumoto}, \citenamefont {Nishiyama},\ and\ \citenamefont
  {Arai}}]{ishida2017amorphous}%
  \BibitemOpen
  \bibfield  {author} {\bibinfo {author} {\bibfnamefont {E.}~\bibnamefont
  {Ishida}}, \bibinfo {author} {\bibfnamefont {K.}~\bibnamefont {Miura}},
  \bibinfo {author} {\bibfnamefont {Y.}~\bibnamefont {Shoji}}, \bibinfo
  {author} {\bibfnamefont {H.}~\bibnamefont {Yokoi}}, \bibinfo {author}
  {\bibfnamefont {T.}~\bibnamefont {Mizumoto}}, \bibinfo {author}
  {\bibfnamefont {N.}~\bibnamefont {Nishiyama}}, \ and\ \bibinfo {author}
  {\bibfnamefont {S.}~\bibnamefont {Arai}},\ }\href@noop {} {\bibfield
  {journal} {\bibinfo  {journal} {Optics express}\ }\textbf {\bibinfo {volume}
  {25}},\ \bibinfo {pages} {452} (\bibinfo {year} {2017})}\BibitemShut
  {NoStop}%
\bibitem [{\citenamefont {Shoji}\ \emph {et~al.}(2008)\citenamefont {Shoji},
  \citenamefont {Mizumoto}, \citenamefont {Yokoi}, \citenamefont {Hsieh},\ and\
  \citenamefont {Osgood~Jr}}]{shoji2008magneto}%
  \BibitemOpen
  \bibfield  {author} {\bibinfo {author} {\bibfnamefont {Y.}~\bibnamefont
  {Shoji}}, \bibinfo {author} {\bibfnamefont {T.}~\bibnamefont {Mizumoto}},
  \bibinfo {author} {\bibfnamefont {H.}~\bibnamefont {Yokoi}}, \bibinfo
  {author} {\bibfnamefont {I.-W.}\ \bibnamefont {Hsieh}}, \ and\ \bibinfo
  {author} {\bibfnamefont {R.~M.}\ \bibnamefont {Osgood~Jr}},\ }\href@noop {}
  {\bibfield  {journal} {\bibinfo  {journal} {Applied physics letters}\
  }\textbf {\bibinfo {volume} {92}},\ \bibinfo {pages} {071117} (\bibinfo
  {year} {2008})}\BibitemShut {NoStop}%
\bibitem [{\citenamefont {Lenz}(1990)}]{lenz1990review}%
  \BibitemOpen
  \bibfield  {author} {\bibinfo {author} {\bibfnamefont {J.~E.}\ \bibnamefont
  {Lenz}},\ }\href@noop {} {\bibfield  {journal} {\bibinfo  {journal}
  {Proceedings of the IEEE}\ }\textbf {\bibinfo {volume} {78}},\ \bibinfo
  {pages} {973} (\bibinfo {year} {1990})}\BibitemShut {NoStop}%
\bibitem [{\citenamefont {Zu}\ \emph {et~al.}(2012)\citenamefont {Zu},
  \citenamefont {Chan}, \citenamefont {Lew}, \citenamefont {Jin}, \citenamefont
  {Zhang}, \citenamefont {Liew}, \citenamefont {Chen}, \citenamefont {Wong},\
  and\ \citenamefont {Dong}}]{zu2012magneto}%
  \BibitemOpen
  \bibfield  {author} {\bibinfo {author} {\bibfnamefont {P.}~\bibnamefont
  {Zu}}, \bibinfo {author} {\bibfnamefont {C.~C.}\ \bibnamefont {Chan}},
  \bibinfo {author} {\bibfnamefont {W.~S.}\ \bibnamefont {Lew}}, \bibinfo
  {author} {\bibfnamefont {Y.}~\bibnamefont {Jin}}, \bibinfo {author}
  {\bibfnamefont {Y.}~\bibnamefont {Zhang}}, \bibinfo {author} {\bibfnamefont
  {H.~F.}\ \bibnamefont {Liew}}, \bibinfo {author} {\bibfnamefont {L.~H.}\
  \bibnamefont {Chen}}, \bibinfo {author} {\bibfnamefont {W.~C.}\ \bibnamefont
  {Wong}}, \ and\ \bibinfo {author} {\bibfnamefont {X.}~\bibnamefont {Dong}},\
  }\href@noop {} {\bibfield  {journal} {\bibinfo  {journal} {Optics letters}\
  }\textbf {\bibinfo {volume} {37}},\ \bibinfo {pages} {398} (\bibinfo {year}
  {2012})}\BibitemShut {NoStop}%
\bibitem [{\citenamefont {Didosyan}\ \emph {et~al.}(2003)\citenamefont
  {Didosyan}, \citenamefont {Hauser}, \citenamefont {Wolfmayr}, \citenamefont
  {Nicolics},\ and\ \citenamefont {Fulmek}}]{didosyan2003magneto}%
  \BibitemOpen
  \bibfield  {author} {\bibinfo {author} {\bibfnamefont {Y.~S.}\ \bibnamefont
  {Didosyan}}, \bibinfo {author} {\bibfnamefont {H.}~\bibnamefont {Hauser}},
  \bibinfo {author} {\bibfnamefont {H.}~\bibnamefont {Wolfmayr}}, \bibinfo
  {author} {\bibfnamefont {J.}~\bibnamefont {Nicolics}}, \ and\ \bibinfo
  {author} {\bibfnamefont {P.}~\bibnamefont {Fulmek}},\ }\href@noop {}
  {\bibfield  {journal} {\bibinfo  {journal} {Sensors and Actuators A:
  Physical}\ }\textbf {\bibinfo {volume} {106}},\ \bibinfo {pages} {168}
  (\bibinfo {year} {2003})}\BibitemShut {NoStop}%
\bibitem [{\citenamefont {Diaz-Valencia}\ \emph {et~al.}(2017)\citenamefont
  {Diaz-Valencia}, \citenamefont {Mej{\'\i}a-Salazar}, \citenamefont
  {Oliveira~Jr}, \citenamefont {Porras-Montenegro},\ and\ \citenamefont
  {Albella}}]{diaz2017enhanced}%
  \BibitemOpen
  \bibfield  {author} {\bibinfo {author} {\bibfnamefont {B.}~\bibnamefont
  {Diaz-Valencia}}, \bibinfo {author} {\bibfnamefont {J.}~\bibnamefont
  {Mej{\'\i}a-Salazar}}, \bibinfo {author} {\bibfnamefont {O.~N.}\ \bibnamefont
  {Oliveira~Jr}}, \bibinfo {author} {\bibfnamefont {N.}~\bibnamefont
  {Porras-Montenegro}}, \ and\ \bibinfo {author} {\bibfnamefont
  {P.}~\bibnamefont {Albella}},\ }\href@noop {} {\bibfield  {journal} {\bibinfo
   {journal} {ACS omega}\ }\textbf {\bibinfo {volume} {2}},\ \bibinfo {pages}
  {7682} (\bibinfo {year} {2017})}\BibitemShut {NoStop}%
\bibitem [{\citenamefont {Sep{\'u}lveda}\ \emph {et~al.}(2006)\citenamefont
  {Sep{\'u}lveda}, \citenamefont {Calle}, \citenamefont {Lechuga},\ and\
  \citenamefont {Armelles}}]{sepulveda2006highly}%
  \BibitemOpen
  \bibfield  {author} {\bibinfo {author} {\bibfnamefont {B.}~\bibnamefont
  {Sep{\'u}lveda}}, \bibinfo {author} {\bibfnamefont {A.}~\bibnamefont
  {Calle}}, \bibinfo {author} {\bibfnamefont {L.~M.}\ \bibnamefont {Lechuga}},
  \ and\ \bibinfo {author} {\bibfnamefont {G.}~\bibnamefont {Armelles}},\
  }\href@noop {} {\bibfield  {journal} {\bibinfo  {journal} {Optics letters}\
  }\textbf {\bibinfo {volume} {31}},\ \bibinfo {pages} {1085} (\bibinfo {year}
  {2006})}\BibitemShut {NoStop}%
\bibitem [{\citenamefont {Park}\ \emph {et~al.}(2009)\citenamefont {Park},
  \citenamefont {Handa},\ and\ \citenamefont {Sandhu}}]{park2009magneto}%
  \BibitemOpen
  \bibfield  {author} {\bibinfo {author} {\bibfnamefont {S.~Y.}\ \bibnamefont
  {Park}}, \bibinfo {author} {\bibfnamefont {H.}~\bibnamefont {Handa}}, \ and\
  \bibinfo {author} {\bibfnamefont {A.}~\bibnamefont {Sandhu}},\ }\href@noop {}
  {\bibfield  {journal} {\bibinfo  {journal} {Nano letters}\ }\textbf {\bibinfo
  {volume} {10}},\ \bibinfo {pages} {446} (\bibinfo {year} {2009})}\BibitemShut
  {NoStop}%
\bibitem [{\citenamefont {Regatos}\ \emph {et~al.}(2011)\citenamefont
  {Regatos}, \citenamefont {Sep{\'u}lveda}, \citenamefont {Fari{\~n}a},
  \citenamefont {Carrascosa},\ and\ \citenamefont
  {Lechuga}}]{regatos2011suitable}%
  \BibitemOpen
  \bibfield  {author} {\bibinfo {author} {\bibfnamefont {D.}~\bibnamefont
  {Regatos}}, \bibinfo {author} {\bibfnamefont {B.}~\bibnamefont
  {Sep{\'u}lveda}}, \bibinfo {author} {\bibfnamefont {D.}~\bibnamefont
  {Fari{\~n}a}}, \bibinfo {author} {\bibfnamefont {L.~G.}\ \bibnamefont
  {Carrascosa}}, \ and\ \bibinfo {author} {\bibfnamefont {L.~M.}\ \bibnamefont
  {Lechuga}},\ }\href@noop {} {\bibfield  {journal} {\bibinfo  {journal}
  {Optics express}\ }\textbf {\bibinfo {volume} {19}},\ \bibinfo {pages} {8336}
  (\bibinfo {year} {2011})}\BibitemShut {NoStop}%
\bibitem [{\citenamefont {Rotondaro}\ \emph {et~al.}(2015)\citenamefont
  {Rotondaro}, \citenamefont {Zhdanov},\ and\ \citenamefont
  {Knize}}]{rotondaro2015generalized}%
  \BibitemOpen
  \bibfield  {author} {\bibinfo {author} {\bibfnamefont {M.~D.}\ \bibnamefont
  {Rotondaro}}, \bibinfo {author} {\bibfnamefont {B.~V.}\ \bibnamefont
  {Zhdanov}}, \ and\ \bibinfo {author} {\bibfnamefont {R.~J.}\ \bibnamefont
  {Knize}},\ }\href@noop {} {\bibfield  {journal} {\bibinfo  {journal} {JOSA
  B}\ }\textbf {\bibinfo {volume} {32}},\ \bibinfo {pages} {2507} (\bibinfo
  {year} {2015})}\BibitemShut {NoStop}%
\bibitem [{\citenamefont {Keaveney}\ \emph {et~al.}(2018)\citenamefont
  {Keaveney}, \citenamefont {Wrathmall}, \citenamefont {Adams},\ and\
  \citenamefont {Hughes}}]{keaveney2018optimized}%
  \BibitemOpen
  \bibfield  {author} {\bibinfo {author} {\bibfnamefont {J.}~\bibnamefont
  {Keaveney}}, \bibinfo {author} {\bibfnamefont {S.~A.}\ \bibnamefont
  {Wrathmall}}, \bibinfo {author} {\bibfnamefont {C.~S.}\ \bibnamefont
  {Adams}}, \ and\ \bibinfo {author} {\bibfnamefont {I.~G.}\ \bibnamefont
  {Hughes}},\ }\href@noop {} {\bibfield  {journal} {\bibinfo  {journal} {Optics
  letters}\ }\textbf {\bibinfo {volume} {43}},\ \bibinfo {pages} {4272}
  (\bibinfo {year} {2018})}\BibitemShut {NoStop}%
\bibitem [{\citenamefont {Bossini}\ \emph {et~al.}(2016)\citenamefont
  {Bossini}, \citenamefont {Belotelov}, \citenamefont {Zvezdin}, \citenamefont
  {Kalish},\ and\ \citenamefont {Kimel}}]{bossini2016}%
  \BibitemOpen
  \bibfield  {author} {\bibinfo {author} {\bibfnamefont {D.}~\bibnamefont
  {Bossini}}, \bibinfo {author} {\bibfnamefont {V.~I.}\ \bibnamefont
  {Belotelov}}, \bibinfo {author} {\bibfnamefont {A.~K.}\ \bibnamefont
  {Zvezdin}}, \bibinfo {author} {\bibfnamefont {A.~N.}\ \bibnamefont {Kalish}},
  \ and\ \bibinfo {author} {\bibfnamefont {A.~V.}\ \bibnamefont {Kimel}},\
  }\href {\doibase 10.1021/acsphotonics.6b00107} {\bibfield  {journal}
  {\bibinfo  {journal} {ACS Photonics}\ }\textbf {\bibinfo {volume} {3}},\
  \bibinfo {pages} {1385} (\bibinfo {year} {2016})}\BibitemShut {NoStop}%
\bibitem [{\citenamefont {Krinchik}\ and\ \citenamefont
  {Artem'ev}(1968)}]{krinchik1968magneto}%
  \BibitemOpen
  \bibfield  {author} {\bibinfo {author} {\bibfnamefont {G.~S.}\ \bibnamefont
  {Krinchik}}\ and\ \bibinfo {author} {\bibfnamefont {V.~A.}\ \bibnamefont
  {Artem'ev}},\ }\href@noop {} {\bibfield  {journal} {\bibinfo  {journal} {Sov.
  Phys. JETP}\ }\textbf {\bibinfo {volume} {26}},\ \bibinfo {pages} {1080}
  (\bibinfo {year} {1968})}\BibitemShut {NoStop}%
\bibitem [{\citenamefont {Martin}\ \emph {et~al.}(1965)\citenamefont {Martin},
  \citenamefont {Neal},\ and\ \citenamefont {Dean}}]{martin1965optical}%
  \BibitemOpen
  \bibfield  {author} {\bibinfo {author} {\bibfnamefont {D.}~\bibnamefont
  {Martin}}, \bibinfo {author} {\bibfnamefont {K.}~\bibnamefont {Neal}}, \ and\
  \bibinfo {author} {\bibfnamefont {T.}~\bibnamefont {Dean}},\ }\href@noop {}
  {\bibfield  {journal} {\bibinfo  {journal} {Proceedings of the Physical
  Society}\ }\textbf {\bibinfo {volume} {86}},\ \bibinfo {pages} {605}
  (\bibinfo {year} {1965})}\BibitemShut {NoStop}%
\bibitem [{\citenamefont {Zvezdin}\ and\ \citenamefont
  {Kotov}(1997)}]{zvezdin1997}%
  \BibitemOpen
  \bibfield  {author} {\bibinfo {author} {\bibfnamefont {A.~K.}\ \bibnamefont
  {Zvezdin}}\ and\ \bibinfo {author} {\bibfnamefont {V.~A.}\ \bibnamefont
  {Kotov}},\ }\href@noop {} {\emph {\bibinfo {title} {Modern Magnetooptics and
  Magnetooptical Materials: Studies in Condensed Matter}}}\ (\bibinfo
  {publisher} {IOP Publishing Ltd},\ \bibinfo {year} {1997})\BibitemShut
  {NoStop}%
\bibitem [{\citenamefont {Grunin}\ \emph {et~al.}(2010)\citenamefont {Grunin},
  \citenamefont {Zhdanov}, \citenamefont {Ezhov}, \citenamefont {Ganshina},\
  and\ \citenamefont {Fedyanin}}]{grunin2010surface}%
  \BibitemOpen
  \bibfield  {author} {\bibinfo {author} {\bibfnamefont {A.}~\bibnamefont
  {Grunin}}, \bibinfo {author} {\bibfnamefont {A.}~\bibnamefont {Zhdanov}},
  \bibinfo {author} {\bibfnamefont {A.}~\bibnamefont {Ezhov}}, \bibinfo
  {author} {\bibfnamefont {E.}~\bibnamefont {Ganshina}}, \ and\ \bibinfo
  {author} {\bibfnamefont {A.}~\bibnamefont {Fedyanin}},\ }\href@noop {}
  {\bibfield  {journal} {\bibinfo  {journal} {Applied Physics Letters}\
  }\textbf {\bibinfo {volume} {97}},\ \bibinfo {pages} {261908} (\bibinfo
  {year} {2010})}\BibitemShut {NoStop}%
\bibitem [{\citenamefont {Maksymov}\ \emph {et~al.}(2014)\citenamefont
  {Maksymov}, \citenamefont {Hutomo},\ and\ \citenamefont
  {Kostylev}}]{maksymov2014transverse}%
  \BibitemOpen
  \bibfield  {author} {\bibinfo {author} {\bibfnamefont {I.~S.}\ \bibnamefont
  {Maksymov}}, \bibinfo {author} {\bibfnamefont {J.}~\bibnamefont {Hutomo}}, \
  and\ \bibinfo {author} {\bibfnamefont {M.}~\bibnamefont {Kostylev}},\
  }\href@noop {} {\bibfield  {journal} {\bibinfo  {journal} {Optics express}\
  }\textbf {\bibinfo {volume} {22}},\ \bibinfo {pages} {8720} (\bibinfo {year}
  {2014})}\BibitemShut {NoStop}%
\bibitem [{\citenamefont {Barsukova}\ \emph {et~al.}(2017)\citenamefont
  {Barsukova}, \citenamefont {Shorokhov}, \citenamefont {Musorin},
  \citenamefont {Neshev}, \citenamefont {Kivshar},\ and\ \citenamefont
  {Fedyanin}}]{barsukova2017magneto}%
  \BibitemOpen
  \bibfield  {author} {\bibinfo {author} {\bibfnamefont {M.~G.}\ \bibnamefont
  {Barsukova}}, \bibinfo {author} {\bibfnamefont {A.~S.}\ \bibnamefont
  {Shorokhov}}, \bibinfo {author} {\bibfnamefont {A.~I.}\ \bibnamefont
  {Musorin}}, \bibinfo {author} {\bibfnamefont {D.~N.}\ \bibnamefont {Neshev}},
  \bibinfo {author} {\bibfnamefont {Y.~S.}\ \bibnamefont {Kivshar}}, \ and\
  \bibinfo {author} {\bibfnamefont {A.~A.}\ \bibnamefont {Fedyanin}},\
  }\href@noop {} {\bibfield  {journal} {\bibinfo  {journal} {ACS Photonics}\
  }\textbf {\bibinfo {volume} {4}},\ \bibinfo {pages} {2390} (\bibinfo {year}
  {2017})}\BibitemShut {NoStop}%
\bibitem [{\citenamefont {Amanollahi}\ and\ \citenamefont
  {Zamani}(2018)}]{amanollahi2018wide}%
  \BibitemOpen
  \bibfield  {author} {\bibinfo {author} {\bibfnamefont {M.}~\bibnamefont
  {Amanollahi}}\ and\ \bibinfo {author} {\bibfnamefont {M.}~\bibnamefont
  {Zamani}},\ }\href@noop {} {\bibfield  {journal} {\bibinfo  {journal} {Optics
  letters}\ }\textbf {\bibinfo {volume} {43}},\ \bibinfo {pages} {4851}
  (\bibinfo {year} {2018})}\BibitemShut {NoStop}%
\bibitem [{\citenamefont {Mukherjee}\ \emph {et~al.}(2019)\citenamefont
  {Mukherjee}, \citenamefont {Ellis}, \citenamefont {Arik}, \citenamefont
  {Taheri}, \citenamefont {Oliverio}, \citenamefont {Fowler}, \citenamefont
  {Tischler}, \citenamefont {Liu}, \citenamefont {Glaser}, \citenamefont
  {Myers-Ward}, \citenamefont {Tedesco}, \citenamefont {Eddy}, \citenamefont
  {Gaskill}, \citenamefont {Zeng}, \citenamefont {Wang},\ and\ \citenamefont
  {Cerne}}]{PhysRevB.99.085440}%
  \BibitemOpen
  \bibfield  {author} {\bibinfo {author} {\bibfnamefont {A.}~\bibnamefont
  {Mukherjee}}, \bibinfo {author} {\bibfnamefont {C.~T.}\ \bibnamefont
  {Ellis}}, \bibinfo {author} {\bibfnamefont {M.~M.}\ \bibnamefont {Arik}},
  \bibinfo {author} {\bibfnamefont {P.}~\bibnamefont {Taheri}}, \bibinfo
  {author} {\bibfnamefont {E.}~\bibnamefont {Oliverio}}, \bibinfo {author}
  {\bibfnamefont {P.}~\bibnamefont {Fowler}}, \bibinfo {author} {\bibfnamefont
  {J.~G.}\ \bibnamefont {Tischler}}, \bibinfo {author} {\bibfnamefont
  {Y.}~\bibnamefont {Liu}}, \bibinfo {author} {\bibfnamefont {E.~R.}\
  \bibnamefont {Glaser}}, \bibinfo {author} {\bibfnamefont {R.~L.}\
  \bibnamefont {Myers-Ward}}, \bibinfo {author} {\bibfnamefont {J.~L.}\
  \bibnamefont {Tedesco}}, \bibinfo {author} {\bibfnamefont {C.~R.}\
  \bibnamefont {Eddy}}, \bibinfo {author} {\bibfnamefont {D.~K.}\ \bibnamefont
  {Gaskill}}, \bibinfo {author} {\bibfnamefont {H.}~\bibnamefont {Zeng}},
  \bibinfo {author} {\bibfnamefont {G.}~\bibnamefont {Wang}}, \ and\ \bibinfo
  {author} {\bibfnamefont {J.}~\bibnamefont {Cerne}},\ }\href {\doibase
  10.1103/PhysRevB.99.085440} {\bibfield  {journal} {\bibinfo  {journal} {Phys.
  Rev. B}\ }\textbf {\bibinfo {volume} {99}},\ \bibinfo {pages} {085440}
  (\bibinfo {year} {2019})}\BibitemShut {NoStop}%
\bibitem [{\citenamefont {Gir{\'o}n-Sedas}\ \emph {et~al.}(2017)\citenamefont
  {Gir{\'o}n-Sedas}, \citenamefont {G{\'o}mez}, \citenamefont {Albella},
  \citenamefont {Mej{\'\i}a-Salazar},\ and\ \citenamefont
  {Oliveira~Jr}}]{giron2017giant}%
  \BibitemOpen
  \bibfield  {author} {\bibinfo {author} {\bibfnamefont {J.}~\bibnamefont
  {Gir{\'o}n-Sedas}}, \bibinfo {author} {\bibfnamefont {F.~R.}\ \bibnamefont
  {G{\'o}mez}}, \bibinfo {author} {\bibfnamefont {P.}~\bibnamefont {Albella}},
  \bibinfo {author} {\bibfnamefont {J.}~\bibnamefont {Mej{\'\i}a-Salazar}}, \
  and\ \bibinfo {author} {\bibfnamefont {O.~N.}\ \bibnamefont {Oliveira~Jr}},\
  }\href@noop {} {\bibfield  {journal} {\bibinfo  {journal} {Physical Review
  B}\ }\textbf {\bibinfo {volume} {96}},\ \bibinfo {pages} {075415} (\bibinfo
  {year} {2017})}\BibitemShut {NoStop}%
\bibitem [{\citenamefont {Nakayama}\ \emph {et~al.}(2017)\citenamefont
  {Nakayama}, \citenamefont {Okano}, \citenamefont {Nozaki},\ and\
  \citenamefont {Watanabe}}]{satoru2017magneto}%
  \BibitemOpen
  \bibfield  {author} {\bibinfo {author} {\bibfnamefont {S.}~\bibnamefont
  {Nakayama}}, \bibinfo {author} {\bibfnamefont {M.}~\bibnamefont {Okano}},
  \bibinfo {author} {\bibfnamefont {Y.}~\bibnamefont {Nozaki}}, \ and\ \bibinfo
  {author} {\bibfnamefont {S.}~\bibnamefont {Watanabe}},\ }\href@noop {}
  {\bibfield  {journal} {\bibinfo  {journal} {AIP Advances}\ }\textbf {\bibinfo
  {volume} {7}},\ \bibinfo {pages} {056802} (\bibinfo {year}
  {2017})}\BibitemShut {NoStop}%
\bibitem [{\citenamefont {Belotelov}\ \emph {et~al.}(2011)\citenamefont
  {Belotelov}, \citenamefont {Akimov}, \citenamefont {Pohl}, \citenamefont
  {Kotov}, \citenamefont {Kasture}, \citenamefont {Vengurlekar}, \citenamefont
  {Gopal}, \citenamefont {Yakovlev}, \citenamefont {Zvezdin},\ and\
  \citenamefont {Bayer}}]{belotelov2011enhanced}%
  \BibitemOpen
  \bibfield  {author} {\bibinfo {author} {\bibfnamefont {V.}~\bibnamefont
  {Belotelov}}, \bibinfo {author} {\bibfnamefont {I.}~\bibnamefont {Akimov}},
  \bibinfo {author} {\bibfnamefont {M.}~\bibnamefont {Pohl}}, \bibinfo {author}
  {\bibfnamefont {V.}~\bibnamefont {Kotov}}, \bibinfo {author} {\bibfnamefont
  {S.}~\bibnamefont {Kasture}}, \bibinfo {author} {\bibfnamefont
  {A.}~\bibnamefont {Vengurlekar}}, \bibinfo {author} {\bibfnamefont {A.~V.}\
  \bibnamefont {Gopal}}, \bibinfo {author} {\bibfnamefont {D.}~\bibnamefont
  {Yakovlev}}, \bibinfo {author} {\bibfnamefont {A.}~\bibnamefont {Zvezdin}}, \
  and\ \bibinfo {author} {\bibfnamefont {M.}~\bibnamefont {Bayer}},\
  }\href@noop {} {\bibfield  {journal} {\bibinfo  {journal} {Nature
  Nanotechnology}\ }\textbf {\bibinfo {volume} {6}},\ \bibinfo {pages} {370}
  (\bibinfo {year} {2011})}\BibitemShut {NoStop}%
\bibitem [{\citenamefont {Akimov}\ \emph {et~al.}(2012)\citenamefont {Akimov},
  \citenamefont {Belotelov}, \citenamefont {Scherbakov}, \citenamefont {Pohl},
  \citenamefont {Kalish}, \citenamefont {Salasyuk}, \citenamefont {Bombeck},
  \citenamefont {Br{\"u}ggemann}, \citenamefont {Akimov}, \citenamefont
  {Dzhioev} \emph {et~al.}}]{akimov2012hybrid}%
  \BibitemOpen
  \bibfield  {author} {\bibinfo {author} {\bibfnamefont {I.~A.}\ \bibnamefont
  {Akimov}}, \bibinfo {author} {\bibfnamefont {V.~I.}\ \bibnamefont
  {Belotelov}}, \bibinfo {author} {\bibfnamefont {A.~V.}\ \bibnamefont
  {Scherbakov}}, \bibinfo {author} {\bibfnamefont {M.}~\bibnamefont {Pohl}},
  \bibinfo {author} {\bibfnamefont {A.~N.}\ \bibnamefont {Kalish}}, \bibinfo
  {author} {\bibfnamefont {A.~S.}\ \bibnamefont {Salasyuk}}, \bibinfo {author}
  {\bibfnamefont {M.}~\bibnamefont {Bombeck}}, \bibinfo {author} {\bibfnamefont
  {C.}~\bibnamefont {Br{\"u}ggemann}}, \bibinfo {author} {\bibfnamefont
  {A.~V.}\ \bibnamefont {Akimov}}, \bibinfo {author} {\bibfnamefont {R.~I.}\
  \bibnamefont {Dzhioev}},  \emph {et~al.},\ }\href@noop {} {\bibfield
  {journal} {\bibinfo  {journal} {JOSA B}\ }\textbf {\bibinfo {volume} {29}},\
  \bibinfo {pages} {A103} (\bibinfo {year} {2012})}\BibitemShut {NoStop}%
\bibitem [{\citenamefont {Gonz{\'a}lez-D{\'\i}az}\ \emph
  {et~al.}(2008)\citenamefont {Gonz{\'a}lez-D{\'\i}az}, \citenamefont
  {Garc{\'\i}a-Mart{\'\i}n}, \citenamefont {Garc{\'\i}a-Mart{\'\i}n},
  \citenamefont {Cebollada}, \citenamefont {Armelles}, \citenamefont
  {Sep{\'u}lveda}, \citenamefont {Alaverdyan},\ and\ \citenamefont
  {K{\"a}ll}}]{gonzalez2008plasmonic}%
  \BibitemOpen
  \bibfield  {author} {\bibinfo {author} {\bibfnamefont {J.~B.}\ \bibnamefont
  {Gonz{\'a}lez-D{\'\i}az}}, \bibinfo {author} {\bibfnamefont {A.}~\bibnamefont
  {Garc{\'\i}a-Mart{\'\i}n}}, \bibinfo {author} {\bibfnamefont {J.~M.}\
  \bibnamefont {Garc{\'\i}a-Mart{\'\i}n}}, \bibinfo {author} {\bibfnamefont
  {A.}~\bibnamefont {Cebollada}}, \bibinfo {author} {\bibfnamefont
  {G.}~\bibnamefont {Armelles}}, \bibinfo {author} {\bibfnamefont
  {B.}~\bibnamefont {Sep{\'u}lveda}}, \bibinfo {author} {\bibfnamefont
  {Y.}~\bibnamefont {Alaverdyan}}, \ and\ \bibinfo {author} {\bibfnamefont
  {M.}~\bibnamefont {K{\"a}ll}},\ }\href@noop {} {\bibfield  {journal}
  {\bibinfo  {journal} {Small}\ }\textbf {\bibinfo {volume} {4}},\ \bibinfo
  {pages} {202} (\bibinfo {year} {2008})}\BibitemShut {NoStop}%
\bibitem [{\citenamefont {Pohl}\ \emph {et~al.}(2013)\citenamefont {Pohl},
  \citenamefont {Kreilkamp}, \citenamefont {Belotelov}, \citenamefont {Akimov},
  \citenamefont {Kalish}, \citenamefont {Khokhlov}, \citenamefont
  {Yallapragada}, \citenamefont {Gopal}, \citenamefont {Nur-E-Alam},
  \citenamefont {Vasiliev} \emph {et~al.}}]{pohl2013tuning}%
  \BibitemOpen
  \bibfield  {author} {\bibinfo {author} {\bibfnamefont {M.}~\bibnamefont
  {Pohl}}, \bibinfo {author} {\bibfnamefont {L.}~\bibnamefont {Kreilkamp}},
  \bibinfo {author} {\bibfnamefont {V.~I.}\ \bibnamefont {Belotelov}}, \bibinfo
  {author} {\bibfnamefont {I.~A.}\ \bibnamefont {Akimov}}, \bibinfo {author}
  {\bibfnamefont {A.~N.}\ \bibnamefont {Kalish}}, \bibinfo {author}
  {\bibfnamefont {N.}~\bibnamefont {Khokhlov}}, \bibinfo {author}
  {\bibfnamefont {V.}~\bibnamefont {Yallapragada}}, \bibinfo {author}
  {\bibfnamefont {A.}~\bibnamefont {Gopal}}, \bibinfo {author} {\bibfnamefont
  {M.}~\bibnamefont {Nur-E-Alam}}, \bibinfo {author} {\bibfnamefont
  {M.}~\bibnamefont {Vasiliev}},  \emph {et~al.},\ }\href@noop {} {\bibfield
  {journal} {\bibinfo  {journal} {New Journal of Physics}\ }\textbf {\bibinfo
  {volume} {15}},\ \bibinfo {pages} {075024} (\bibinfo {year}
  {2013})}\BibitemShut {NoStop}%
\bibitem [{\citenamefont {Clavero}\ \emph {et~al.}(2010)\citenamefont
  {Clavero}, \citenamefont {Yang}, \citenamefont {Skuza},\ and\ \citenamefont
  {Lukaszew}}]{clavero2010magnetic}%
  \BibitemOpen
  \bibfield  {author} {\bibinfo {author} {\bibfnamefont {C.}~\bibnamefont
  {Clavero}}, \bibinfo {author} {\bibfnamefont {K.}~\bibnamefont {Yang}},
  \bibinfo {author} {\bibfnamefont {J.}~\bibnamefont {Skuza}}, \ and\ \bibinfo
  {author} {\bibfnamefont {R.}~\bibnamefont {Lukaszew}},\ }\href@noop {}
  {\bibfield  {journal} {\bibinfo  {journal} {Optics letters}\ }\textbf
  {\bibinfo {volume} {35}},\ \bibinfo {pages} {1557} (\bibinfo {year}
  {2010})}\BibitemShut {NoStop}%
\bibitem [{\citenamefont {Newman}\ \emph {et~al.}(2008)\citenamefont {Newman},
  \citenamefont {Wears}, \citenamefont {Matelon},\ and\ \citenamefont
  {Hooper}}]{newman2008magneto}%
  \BibitemOpen
  \bibfield  {author} {\bibinfo {author} {\bibfnamefont {D.}~\bibnamefont
  {Newman}}, \bibinfo {author} {\bibfnamefont {M.}~\bibnamefont {Wears}},
  \bibinfo {author} {\bibfnamefont {R.}~\bibnamefont {Matelon}}, \ and\
  \bibinfo {author} {\bibfnamefont {I.}~\bibnamefont {Hooper}},\ }\href@noop {}
  {\bibfield  {journal} {\bibinfo  {journal} {Journal of Physics: Condensed
  Matter}\ }\textbf {\bibinfo {volume} {20}},\ \bibinfo {pages} {345230}
  (\bibinfo {year} {2008})}\BibitemShut {NoStop}%
\bibitem [{\citenamefont {Dyakov}\ \emph
  {et~al.}(2018{\natexlab{a}})\citenamefont {Dyakov}, \citenamefont {Spitzer},
  \citenamefont {Akimov}, \citenamefont {Yavsin}, \citenamefont {Pavlov},
  \citenamefont {Verbin}, \citenamefont {Tikhodeev}, \citenamefont {Gippius},
  \citenamefont {Pevtsov},\ and\ \citenamefont {Bayer}}]{dyakov2018transverse}%
  \BibitemOpen
  \bibfield  {author} {\bibinfo {author} {\bibfnamefont {S.}~\bibnamefont
  {Dyakov}}, \bibinfo {author} {\bibfnamefont {F.}~\bibnamefont {Spitzer}},
  \bibinfo {author} {\bibfnamefont {I.}~\bibnamefont {Akimov}}, \bibinfo
  {author} {\bibfnamefont {D.}~\bibnamefont {Yavsin}}, \bibinfo {author}
  {\bibfnamefont {S.}~\bibnamefont {Pavlov}}, \bibinfo {author} {\bibfnamefont
  {S.}~\bibnamefont {Verbin}}, \bibinfo {author} {\bibfnamefont
  {S.}~\bibnamefont {Tikhodeev}}, \bibinfo {author} {\bibfnamefont
  {N.}~\bibnamefont {Gippius}}, \bibinfo {author} {\bibfnamefont
  {A.}~\bibnamefont {Pevtsov}}, \ and\ \bibinfo {author} {\bibfnamefont
  {M.}~\bibnamefont {Bayer}},\ }\href@noop {} {\bibfield  {journal} {\bibinfo
  {journal} {Semiconductors}\ }\textbf {\bibinfo {volume} {52}},\ \bibinfo
  {pages} {1857} (\bibinfo {year} {2018}{\natexlab{a}})}\BibitemShut {NoStop}%
\bibitem [{\citenamefont {Borovkova}\ \emph {et~al.}(2018)\citenamefont
  {Borovkova}, \citenamefont {Hashim}, \citenamefont {Kozhaev}, \citenamefont
  {Dagesyan}, \citenamefont {Chakravarty}, \citenamefont {Levy},\ and\
  \citenamefont {Belotelov}}]{borovkova2018tmoke}%
  \BibitemOpen
  \bibfield  {author} {\bibinfo {author} {\bibfnamefont {O.}~\bibnamefont
  {Borovkova}}, \bibinfo {author} {\bibfnamefont {H.}~\bibnamefont {Hashim}},
  \bibinfo {author} {\bibfnamefont {M.}~\bibnamefont {Kozhaev}}, \bibinfo
  {author} {\bibfnamefont {S.}~\bibnamefont {Dagesyan}}, \bibinfo {author}
  {\bibfnamefont {A.}~\bibnamefont {Chakravarty}}, \bibinfo {author}
  {\bibfnamefont {M.}~\bibnamefont {Levy}}, \ and\ \bibinfo {author}
  {\bibfnamefont {V.}~\bibnamefont {Belotelov}},\ }\href@noop {} {\bibfield
  {journal} {\bibinfo  {journal} {Applied Physics Letters}\ }\textbf {\bibinfo
  {volume} {112}},\ \bibinfo {pages} {063101} (\bibinfo {year}
  {2018})}\BibitemShut {NoStop}%
\bibitem [{\citenamefont {Maksymov}(2016)}]{maksymov2016magneto}%
  \BibitemOpen
  \bibfield  {author} {\bibinfo {author} {\bibfnamefont {I.~S.}\ \bibnamefont
  {Maksymov}},\ }\href@noop {} {\bibfield  {journal} {\bibinfo  {journal}
  {Reviews in Physics}\ }\textbf {\bibinfo {volume} {1}},\ \bibinfo {pages}
  {36} (\bibinfo {year} {2016})}\BibitemShut {NoStop}%
\bibitem [{\citenamefont {Chen}\ \emph {et~al.}(2011)\citenamefont {Chen},
  \citenamefont {Albella}, \citenamefont {Pirzadeh}, \citenamefont
  {Alonso-Gonz{\'a}lez}, \citenamefont {Huth}, \citenamefont {Bonetti},
  \citenamefont {Bonanni}, \citenamefont {{\AA}kerman}, \citenamefont
  {Nogu{\'e}s}, \citenamefont {Vavassori} \emph {et~al.}}]{chen2011plasmonic}%
  \BibitemOpen
  \bibfield  {author} {\bibinfo {author} {\bibfnamefont {J.}~\bibnamefont
  {Chen}}, \bibinfo {author} {\bibfnamefont {P.}~\bibnamefont {Albella}},
  \bibinfo {author} {\bibfnamefont {Z.}~\bibnamefont {Pirzadeh}}, \bibinfo
  {author} {\bibfnamefont {P.}~\bibnamefont {Alonso-Gonz{\'a}lez}}, \bibinfo
  {author} {\bibfnamefont {F.}~\bibnamefont {Huth}}, \bibinfo {author}
  {\bibfnamefont {S.}~\bibnamefont {Bonetti}}, \bibinfo {author} {\bibfnamefont
  {V.}~\bibnamefont {Bonanni}}, \bibinfo {author} {\bibfnamefont
  {J.}~\bibnamefont {{\AA}kerman}}, \bibinfo {author} {\bibfnamefont
  {J.}~\bibnamefont {Nogu{\'e}s}}, \bibinfo {author} {\bibfnamefont
  {P.}~\bibnamefont {Vavassori}},  \emph {et~al.},\ }\href@noop {} {\bibfield
  {journal} {\bibinfo  {journal} {Small}\ }\textbf {\bibinfo {volume} {7}},\
  \bibinfo {pages} {2341} (\bibinfo {year} {2011})}\BibitemShut {NoStop}%
\bibitem [{\citenamefont {Valente}\ \emph {et~al.}(2015)\citenamefont
  {Valente}, \citenamefont {Ou}, \citenamefont {Plum}, \citenamefont {Youngs},\
  and\ \citenamefont {Zheludev}}]{valente2015magneto}%
  \BibitemOpen
  \bibfield  {author} {\bibinfo {author} {\bibfnamefont {J.}~\bibnamefont
  {Valente}}, \bibinfo {author} {\bibfnamefont {J.-Y.}\ \bibnamefont {Ou}},
  \bibinfo {author} {\bibfnamefont {E.}~\bibnamefont {Plum}}, \bibinfo {author}
  {\bibfnamefont {I.~J.}\ \bibnamefont {Youngs}}, \ and\ \bibinfo {author}
  {\bibfnamefont {N.~I.}\ \bibnamefont {Zheludev}},\ }\href@noop {} {\bibfield
  {journal} {\bibinfo  {journal} {Nature communications}\ }\textbf {\bibinfo
  {volume} {6}},\ \bibinfo {pages} {7021} (\bibinfo {year} {2015})}\BibitemShut
  {NoStop}%
\bibitem [{\citenamefont {Loughran}\ \emph {et~al.}(2018)\citenamefont
  {Loughran}, \citenamefont {Keatley}, \citenamefont {Hendry}, \citenamefont
  {Barnes},\ and\ \citenamefont {Hicken}}]{loughran2018enhancing}%
  \BibitemOpen
  \bibfield  {author} {\bibinfo {author} {\bibfnamefont {T.}~\bibnamefont
  {Loughran}}, \bibinfo {author} {\bibfnamefont {P.}~\bibnamefont {Keatley}},
  \bibinfo {author} {\bibfnamefont {E.}~\bibnamefont {Hendry}}, \bibinfo
  {author} {\bibfnamefont {W.}~\bibnamefont {Barnes}}, \ and\ \bibinfo {author}
  {\bibfnamefont {R.}~\bibnamefont {Hicken}},\ }\href@noop {} {\bibfield
  {journal} {\bibinfo  {journal} {Optics express}\ }\textbf {\bibinfo {volume}
  {26}},\ \bibinfo {pages} {4738} (\bibinfo {year} {2018})}\BibitemShut
  {NoStop}%
\bibitem [{\citenamefont {Kreilkamp}\ \emph {et~al.}(2013)\citenamefont
  {Kreilkamp}, \citenamefont {Belotelov}, \citenamefont {Chin}, \citenamefont
  {Neutzner}, \citenamefont {Dregely}, \citenamefont {Wehlus}, \citenamefont
  {Akimov}, \citenamefont {Bayer}, \citenamefont {Stritzker},\ and\
  \citenamefont {Giessen}}]{kreilkamp2013waveguide}%
  \BibitemOpen
  \bibfield  {author} {\bibinfo {author} {\bibfnamefont {L.~E.}\ \bibnamefont
  {Kreilkamp}}, \bibinfo {author} {\bibfnamefont {V.~I.}\ \bibnamefont
  {Belotelov}}, \bibinfo {author} {\bibfnamefont {J.~Y.}\ \bibnamefont {Chin}},
  \bibinfo {author} {\bibfnamefont {S.}~\bibnamefont {Neutzner}}, \bibinfo
  {author} {\bibfnamefont {D.}~\bibnamefont {Dregely}}, \bibinfo {author}
  {\bibfnamefont {T.}~\bibnamefont {Wehlus}}, \bibinfo {author} {\bibfnamefont
  {I.~A.}\ \bibnamefont {Akimov}}, \bibinfo {author} {\bibfnamefont
  {M.}~\bibnamefont {Bayer}}, \bibinfo {author} {\bibfnamefont
  {B.}~\bibnamefont {Stritzker}}, \ and\ \bibinfo {author} {\bibfnamefont
  {H.}~\bibnamefont {Giessen}},\ }\href@noop {} {\bibfield  {journal} {\bibinfo
   {journal} {Physical Review X}\ }\textbf {\bibinfo {volume} {3}},\ \bibinfo
  {pages} {041019} (\bibinfo {year} {2013})}\BibitemShut {NoStop}%
\bibitem [{\citenamefont {Christ}\ \emph {et~al.}(2003)\citenamefont {Christ},
  \citenamefont {Tikhodeev}, \citenamefont {Gippius}, \citenamefont {Kuhl},\
  and\ \citenamefont {Giessen}}]{Christ2003b}%
  \BibitemOpen
  \bibfield  {author} {\bibinfo {author} {\bibfnamefont {A.}~\bibnamefont
  {Christ}}, \bibinfo {author} {\bibfnamefont {S.~G.}\ \bibnamefont
  {Tikhodeev}}, \bibinfo {author} {\bibfnamefont {N.~A.}\ \bibnamefont
  {Gippius}}, \bibinfo {author} {\bibfnamefont {J.}~\bibnamefont {Kuhl}}, \
  and\ \bibinfo {author} {\bibfnamefont {H.}~\bibnamefont {Giessen}},\ }\href
  {https://link.aps.org/doi/10.1103/PhysRevLett.91.183901} {\bibfield
  {journal} {\bibinfo  {journal} {Phys. Rev. Lett.}\ }\textbf {\bibinfo
  {volume} {91}},\ \bibinfo {pages} {183901} (\bibinfo {year}
  {2003})}\BibitemShut {NoStop}%
\bibitem [{\citenamefont {Dyakov}\ \emph {et~al.}(2016)\citenamefont {Dyakov},
  \citenamefont {Zhigunov}, \citenamefont {Marinins}, \citenamefont
  {Shcherbakov}, \citenamefont {Fedyanin}, \citenamefont {Vorontsov},
  \citenamefont {Kashkarov}, \citenamefont {Popov}, \citenamefont {Qiu},
  \citenamefont {Zacharias}, \citenamefont {Tikhodeev},\ and\ \citenamefont
  {Gippius}}]{PhysRevB.93.205413}%
  \BibitemOpen
  \bibfield  {author} {\bibinfo {author} {\bibfnamefont {S.~A.}\ \bibnamefont
  {Dyakov}}, \bibinfo {author} {\bibfnamefont {D.~M.}\ \bibnamefont
  {Zhigunov}}, \bibinfo {author} {\bibfnamefont {A.}~\bibnamefont {Marinins}},
  \bibinfo {author} {\bibfnamefont {M.~R.}\ \bibnamefont {Shcherbakov}},
  \bibinfo {author} {\bibfnamefont {A.~A.}\ \bibnamefont {Fedyanin}}, \bibinfo
  {author} {\bibfnamefont {A.~S.}\ \bibnamefont {Vorontsov}}, \bibinfo {author}
  {\bibfnamefont {P.~K.}\ \bibnamefont {Kashkarov}}, \bibinfo {author}
  {\bibfnamefont {S.}~\bibnamefont {Popov}}, \bibinfo {author} {\bibfnamefont
  {M.}~\bibnamefont {Qiu}}, \bibinfo {author} {\bibfnamefont {M.}~\bibnamefont
  {Zacharias}}, \bibinfo {author} {\bibfnamefont {S.~G.}\ \bibnamefont
  {Tikhodeev}}, \ and\ \bibinfo {author} {\bibfnamefont {N.~A.}\ \bibnamefont
  {Gippius}},\ }\href {http://link.aps.org/doi/10.1103/PhysRevB.93.205413}
  {\bibfield  {journal} {\bibinfo  {journal} {Phys. Rev. B}\ }\textbf {\bibinfo
  {volume} {93}},\ \bibinfo {pages} {205413} (\bibinfo {year}
  {2016})}\BibitemShut {NoStop}%
\bibitem [{\citenamefont {Melekh}\ \emph {et~al.}(2016)\citenamefont {Melekh},
  \citenamefont {Kurdyukov}, \citenamefont {Yavsin}, \citenamefont {Kozhevin},
  \citenamefont {Gurevich}, \citenamefont {Gastev}, \citenamefont {Volkov},
  \citenamefont {Sitnikova}, \citenamefont {Yagovkina},\ and\ \citenamefont
  {Pevtsov}}]{melekh2016nanostructured}%
  \BibitemOpen
  \bibfield  {author} {\bibinfo {author} {\bibfnamefont {B.}~\bibnamefont
  {Melekh}}, \bibinfo {author} {\bibfnamefont {D.}~\bibnamefont {Kurdyukov}},
  \bibinfo {author} {\bibfnamefont {D.}~\bibnamefont {Yavsin}}, \bibinfo
  {author} {\bibfnamefont {V.}~\bibnamefont {Kozhevin}}, \bibinfo {author}
  {\bibfnamefont {S.}~\bibnamefont {Gurevich}}, \bibinfo {author}
  {\bibfnamefont {S.}~\bibnamefont {Gastev}}, \bibinfo {author} {\bibfnamefont
  {M.}~\bibnamefont {Volkov}}, \bibinfo {author} {\bibfnamefont
  {A.}~\bibnamefont {Sitnikova}}, \bibinfo {author} {\bibfnamefont
  {M.}~\bibnamefont {Yagovkina}}, \ and\ \bibinfo {author} {\bibfnamefont
  {A.}~\bibnamefont {Pevtsov}},\ }\href@noop {} {\bibfield  {journal} {\bibinfo
   {journal} {Technical Physics Letters}\ }\textbf {\bibinfo {volume} {42}},\
  \bibinfo {pages} {1005} (\bibinfo {year} {2016})}\BibitemShut {NoStop}%
\bibitem [{\citenamefont {Fontijn}\ \emph {et~al.}(1997)\citenamefont
  {Fontijn}, \citenamefont {Van~der Zaag}, \citenamefont {Devillers},
  \citenamefont {Brabers},\ and\ \citenamefont
  {Metselaar}}]{fontijn1997optical}%
  \BibitemOpen
  \bibfield  {author} {\bibinfo {author} {\bibfnamefont {W.}~\bibnamefont
  {Fontijn}}, \bibinfo {author} {\bibfnamefont {P.}~\bibnamefont {Van~der
  Zaag}}, \bibinfo {author} {\bibfnamefont {M.}~\bibnamefont {Devillers}},
  \bibinfo {author} {\bibfnamefont {V.}~\bibnamefont {Brabers}}, \ and\
  \bibinfo {author} {\bibfnamefont {R.}~\bibnamefont {Metselaar}},\ }\href@noop
  {} {\bibfield  {journal} {\bibinfo  {journal} {Physical Review B}\ }\textbf
  {\bibinfo {volume} {56}},\ \bibinfo {pages} {5432} (\bibinfo {year}
  {1997})}\BibitemShut {NoStop}%
\bibitem [{\citenamefont {Bobo}\ \emph {et~al.}(2001)\citenamefont {Bobo},
  \citenamefont {Basso}, \citenamefont {Snoeck}, \citenamefont {Gatel},
  \citenamefont {Hrabovsky}, \citenamefont {Gauffier}, \citenamefont {Ressier},
  \citenamefont {Mamy}, \citenamefont {Visnovsky}, \citenamefont {Hamrle} \emph
  {et~al.}}]{bobo2001magnetic}%
  \BibitemOpen
  \bibfield  {author} {\bibinfo {author} {\bibfnamefont {J.}~\bibnamefont
  {Bobo}}, \bibinfo {author} {\bibfnamefont {D.}~\bibnamefont {Basso}},
  \bibinfo {author} {\bibfnamefont {E.}~\bibnamefont {Snoeck}}, \bibinfo
  {author} {\bibfnamefont {C.}~\bibnamefont {Gatel}}, \bibinfo {author}
  {\bibfnamefont {D.}~\bibnamefont {Hrabovsky}}, \bibinfo {author}
  {\bibfnamefont {J.}~\bibnamefont {Gauffier}}, \bibinfo {author}
  {\bibfnamefont {L.}~\bibnamefont {Ressier}}, \bibinfo {author} {\bibfnamefont
  {R.}~\bibnamefont {Mamy}}, \bibinfo {author} {\bibfnamefont {S.}~\bibnamefont
  {Visnovsky}}, \bibinfo {author} {\bibfnamefont {J.}~\bibnamefont {Hamrle}},
  \emph {et~al.},\ }\href@noop {} {\bibfield  {journal} {\bibinfo  {journal}
  {The European Physical Journal B-Condensed Matter and Complex Systems}\
  }\textbf {\bibinfo {volume} {24}},\ \bibinfo {pages} {43} (\bibinfo {year}
  {2001})}\BibitemShut {NoStop}%
\bibitem [{\citenamefont {Sehmi}\ \emph {et~al.}(2017)\citenamefont {Sehmi},
  \citenamefont {Langbein},\ and\ \citenamefont
  {Muljarov}}]{sehmi2017optimizing}%
  \BibitemOpen
  \bibfield  {author} {\bibinfo {author} {\bibfnamefont {H.}~\bibnamefont
  {Sehmi}}, \bibinfo {author} {\bibfnamefont {W.}~\bibnamefont {Langbein}}, \
  and\ \bibinfo {author} {\bibfnamefont {E.}~\bibnamefont {Muljarov}},\
  }\href@noop {} {\bibfield  {journal} {\bibinfo  {journal} {Physical Review
  B}\ }\textbf {\bibinfo {volume} {95}},\ \bibinfo {pages} {115444} (\bibinfo
  {year} {2017})}\BibitemShut {NoStop}%
\bibitem [{\citenamefont {Johnson}\ and\ \citenamefont
  {Christy}(1972)}]{johnson1972optical}%
  \BibitemOpen
  \bibfield  {author} {\bibinfo {author} {\bibfnamefont {P.~B.}\ \bibnamefont
  {Johnson}}\ and\ \bibinfo {author} {\bibfnamefont {R.-W.}\ \bibnamefont
  {Christy}},\ }\href@noop {} {\bibfield  {journal} {\bibinfo  {journal}
  {Physical review B}\ }\textbf {\bibinfo {volume} {6}},\ \bibinfo {pages}
  {4370} (\bibinfo {year} {1972})}\BibitemShut {NoStop}%
\bibitem [{\citenamefont {Tikhodeev}\ \emph {et~al.}(2002)\citenamefont
  {Tikhodeev}, \citenamefont {Yablonskii}, \citenamefont {Muljarov},
  \citenamefont {Gippius},\ and\ \citenamefont {Ishihara}}]{Tikhodeev2002b}%
  \BibitemOpen
  \bibfield  {author} {\bibinfo {author} {\bibfnamefont {S.~G.}\ \bibnamefont
  {Tikhodeev}}, \bibinfo {author} {\bibfnamefont {A.~L.}\ \bibnamefont
  {Yablonskii}}, \bibinfo {author} {\bibfnamefont {E.~A.}\ \bibnamefont
  {Muljarov}}, \bibinfo {author} {\bibfnamefont {N.~A.}\ \bibnamefont
  {Gippius}}, \ and\ \bibinfo {author} {\bibfnamefont {T.}~\bibnamefont
  {Ishihara}},\ }\href {https://link.aps.org/doi/10.1103/PhysRevB.66.045102}
  {\bibfield  {journal} {\bibinfo  {journal} {Phys. Rev. B}\ }\textbf {\bibinfo
  {volume} {66}},\ \bibinfo {pages} {045102} (\bibinfo {year}
  {2002})}\BibitemShut {NoStop}%
\bibitem [{\citenamefont {Moharam}\ \emph {et~al.}(1995)\citenamefont
  {Moharam}, \citenamefont {Gaylord}, \citenamefont {Grann},\ and\
  \citenamefont {Pommet}}]{moharam1995formulation}%
  \BibitemOpen
  \bibfield  {author} {\bibinfo {author} {\bibfnamefont {M.}~\bibnamefont
  {Moharam}}, \bibinfo {author} {\bibfnamefont {T.}~\bibnamefont {Gaylord}},
  \bibinfo {author} {\bibfnamefont {E.~B.}\ \bibnamefont {Grann}}, \ and\
  \bibinfo {author} {\bibfnamefont {D.~A.}\ \bibnamefont {Pommet}},\ }\href
  {https://doi.org/10.1364/JOSAA.12.001068} {\bibfield  {journal} {\bibinfo
  {journal} {JOSA a}\ }\textbf {\bibinfo {volume} {12}},\ \bibinfo {pages}
  {1068} (\bibinfo {year} {1995})}\BibitemShut {NoStop}%
\bibitem [{\citenamefont {Li}(1996)}]{li1996use}%
  \BibitemOpen
  \bibfield  {author} {\bibinfo {author} {\bibfnamefont {L.}~\bibnamefont
  {Li}},\ }\href {https://doi.org/10.1364/JOSAA.13.001870} {\bibfield
  {journal} {\bibinfo  {journal} {J. Opt. Soc. Am. A}\ }\textbf {\bibinfo
  {volume} {13}},\ \bibinfo {pages} {1870} (\bibinfo {year}
  {1996})}\BibitemShut {NoStop}%
\bibitem [{\citenamefont {Gippius}\ \emph {et~al.}(2005)\citenamefont
  {Gippius}, \citenamefont {Tikhodeev},\ and\ \citenamefont
  {Ishihara}}]{Gippius2005c}%
  \BibitemOpen
  \bibfield  {author} {\bibinfo {author} {\bibfnamefont {N.}~\bibnamefont
  {Gippius}}, \bibinfo {author} {\bibfnamefont {S.}~\bibnamefont {Tikhodeev}},
  \ and\ \bibinfo {author} {\bibfnamefont {T.}~\bibnamefont {Ishihara}},\
  }\href {https://link.aps.org/doi/10.1103/PhysRevB.72.045138} {\bibfield
  {journal} {\bibinfo  {journal} {Phys. Rev. B}\ }\textbf {\bibinfo {volume}
  {72}},\ \bibinfo {pages} {045138} (\bibinfo {year} {2005})}\BibitemShut
  {NoStop}%
\bibitem [{\citenamefont {Dyakov}\ \emph {et~al.}(2017)\citenamefont {Dyakov},
  \citenamefont {Gippius}, \citenamefont {Voronov}, \citenamefont {Yakovlev},
  \citenamefont {Pevtsov}, \citenamefont {Akimov},\ and\ \citenamefont
  {Tikhodeev}}]{dyakov2017quasiguided}%
  \BibitemOpen
  \bibfield  {author} {\bibinfo {author} {\bibfnamefont {S.~A.}\ \bibnamefont
  {Dyakov}}, \bibinfo {author} {\bibfnamefont {N.~A.}\ \bibnamefont {Gippius}},
  \bibinfo {author} {\bibfnamefont {M.~M.}\ \bibnamefont {Voronov}}, \bibinfo
  {author} {\bibfnamefont {S.~A.}\ \bibnamefont {Yakovlev}}, \bibinfo {author}
  {\bibfnamefont {A.~B.}\ \bibnamefont {Pevtsov}}, \bibinfo {author}
  {\bibfnamefont {I.~A.}\ \bibnamefont {Akimov}}, \ and\ \bibinfo {author}
  {\bibfnamefont {S.~G.}\ \bibnamefont {Tikhodeev}},\ }\href@noop {} {\bibfield
   {journal} {\bibinfo  {journal} {Phys. Rev. B}\ }\textbf {\bibinfo {volume}
  {96}},\ \bibinfo {pages} {045426} (\bibinfo {year} {2017})}\BibitemShut
  {NoStop}%
\bibitem [{\citenamefont {Sakoda}(2001)}]{Sakoda2001a}%
  \BibitemOpen
  \bibfield  {author} {\bibinfo {author} {\bibfnamefont {K.}~\bibnamefont
  {Sakoda}},\ }\href@noop {} {\emph {\bibinfo {title} {Optical Properties of
  Photonic Crystals}}}\ (\bibinfo  {publisher} {Springer},\ \bibinfo {year}
  {2001})\BibitemShut {NoStop}%
\bibitem [{\citenamefont {Dyakov}\ \emph
  {et~al.}(2018{\natexlab{b}})\citenamefont {Dyakov}, \citenamefont
  {Semenenko}, \citenamefont {Gippius},\ and\ \citenamefont
  {Tikhodeev}}]{dyakov2018magnetic}%
  \BibitemOpen
  \bibfield  {author} {\bibinfo {author} {\bibfnamefont {S.}~\bibnamefont
  {Dyakov}}, \bibinfo {author} {\bibfnamefont {V.}~\bibnamefont {Semenenko}},
  \bibinfo {author} {\bibfnamefont {N.}~\bibnamefont {Gippius}}, \ and\
  \bibinfo {author} {\bibfnamefont {S.}~\bibnamefont {Tikhodeev}},\ }\href@noop
  {} {\bibfield  {journal} {\bibinfo  {journal} {Physical Review B}\ }\textbf
  {\bibinfo {volume} {98}},\ \bibinfo {pages} {235416} (\bibinfo {year}
  {2018}{\natexlab{b}})}\BibitemShut {NoStop}%
\bibitem [{\citenamefont {Christ}\ \emph {et~al.}(2004)\citenamefont {Christ},
  \citenamefont {Zentgraf}, \citenamefont {Kuhl}, \citenamefont {Tikhodeev},
  \citenamefont {Gippius},\ and\ \citenamefont {Giessen}}]{Christ2004}%
  \BibitemOpen
  \bibfield  {author} {\bibinfo {author} {\bibfnamefont {A.}~\bibnamefont
  {Christ}}, \bibinfo {author} {\bibfnamefont {T.}~\bibnamefont {Zentgraf}},
  \bibinfo {author} {\bibfnamefont {J.}~\bibnamefont {Kuhl}}, \bibinfo {author}
  {\bibfnamefont {S.~G.}\ \bibnamefont {Tikhodeev}}, \bibinfo {author}
  {\bibfnamefont {N.~A.}\ \bibnamefont {Gippius}}, \ and\ \bibinfo {author}
  {\bibfnamefont {H.}~\bibnamefont {Giessen}},\ }\href
  {https://link.aps.org/doi/10.1103/PhysRevB.70.125113} {\bibfield  {journal}
  {\bibinfo  {journal} {Phys. Rev. B}\ }\textbf {\bibinfo {volume} {70}},\
  \bibinfo {pages} {125113} (\bibinfo {year} {2004})}\BibitemShut {NoStop}%
\bibitem [{Note1()}]{Note1}%
  \BibitemOpen
  \bibinfo {note} {Please note that the model of sample C used in simulations
  includes air trenches with 40 nm depth and 180 nm width.}\BibitemShut {Stop}%
\bibitem [{\citenamefont {Tikhodeev}\ \emph {et~al.}(2005)\citenamefont
  {Tikhodeev}, \citenamefont {Gippius}, \citenamefont {Christ}, \citenamefont
  {Zentgraf}, \citenamefont {Kuhl},\ and\ \citenamefont
  {Giessen}}]{tikhodeev2005waveguide}%
  \BibitemOpen
  \bibfield  {author} {\bibinfo {author} {\bibfnamefont {S.}~\bibnamefont
  {Tikhodeev}}, \bibinfo {author} {\bibfnamefont {N.}~\bibnamefont {Gippius}},
  \bibinfo {author} {\bibfnamefont {A.}~\bibnamefont {Christ}}, \bibinfo
  {author} {\bibfnamefont {T.}~\bibnamefont {Zentgraf}}, \bibinfo {author}
  {\bibfnamefont {J.}~\bibnamefont {Kuhl}}, \ and\ \bibinfo {author}
  {\bibfnamefont {H.}~\bibnamefont {Giessen}},\ }\href@noop {} {\bibfield
  {journal} {\bibinfo  {journal} {physica status solidi (c)}\ }\textbf
  {\bibinfo {volume} {2}},\ \bibinfo {pages} {795} (\bibinfo {year}
  {2005})}\BibitemShut {NoStop}%
\bibitem [{\citenamefont {Chikazumi}\ and\ \citenamefont
  {Graham}(2009)}]{chikazumi2009physics}%
  \BibitemOpen
  \bibfield  {author} {\bibinfo {author} {\bibfnamefont {S.}~\bibnamefont
  {Chikazumi}}\ and\ \bibinfo {author} {\bibfnamefont {C.~D.}\ \bibnamefont
  {Graham}},\ }\href@noop {} {\emph {\bibinfo {title} {Physics of
  Ferromagnetism 2e}}},\ Vol.~\bibinfo {volume} {94}\ (\bibinfo  {publisher}
  {Oxford University Press on Demand},\ \bibinfo {year} {2009})\BibitemShut
  {NoStop}%
\bibitem [{\citenamefont {Richard}\ \emph {et~al.}(2005)\citenamefont
  {Richard}, \citenamefont {Kasprzak}, \citenamefont {Romestain}, \citenamefont
  {Andr{\'e}},\ and\ \citenamefont {Dang}}]{richard2005spontaneous}%
  \BibitemOpen
  \bibfield  {author} {\bibinfo {author} {\bibfnamefont {M.}~\bibnamefont
  {Richard}}, \bibinfo {author} {\bibfnamefont {J.}~\bibnamefont {Kasprzak}},
  \bibinfo {author} {\bibfnamefont {R.}~\bibnamefont {Romestain}}, \bibinfo
  {author} {\bibfnamefont {R.}~\bibnamefont {Andr{\'e}}}, \ and\ \bibinfo
  {author} {\bibfnamefont {L.~S.}\ \bibnamefont {Dang}},\ }\href@noop {}
  {\bibfield  {journal} {\bibinfo  {journal} {Physical review letters}\
  }\textbf {\bibinfo {volume} {94}},\ \bibinfo {pages} {187401} (\bibinfo
  {year} {2005})}\BibitemShut {NoStop}%
\end{thebibliography}
 %
 
\end{document}